\documentclass{article}

\usepackage{booktabs}
\usepackage{graphics}
\usepackage{graphicx}
\usepackage{listings}
\usepackage{nameref}
\usepackage{algorithm}
\usepackage{algpseudocode}
\algrenewcommand\algorithmicrequire{\textbf{Input:}}
\algrenewcommand\algorithmicensure{\textbf{Output:}}
\usepackage{bm}
\usepackage{amsmath}
\usepackage{amssymb}
\usepackage{caption} 
\usepackage{multirow}

\usepackage{arxiv}

\usepackage[utf8]{inputenc} 
\usepackage[T1]{fontenc}    
\usepackage{hyperref}       
\usepackage{url}            
\usepackage{booktabs}       
\usepackage{amsfonts}       
\usepackage{nicefrac}       
\usepackage{microtype}      
\usepackage{lipsum}
\usepackage{graphicx}
\graphicspath{ {./images/} }

\title{CHERRY: a Computational metHod for accuratE pRediction of virus-pRokarYotic interactions using a graph encoder-decoder model}

\author{
 Jiayu Shang \\
  Dept. of Electrical Engineering\\
  City University of Hong Kong\\
  Kowloon, Hong Kong SAR, China\\
  \texttt{jyshang2-c@my.cityu.edu.hk} \\
  \And
 Yanni Sun \\
  Dept. of Electrical Engineering\\
  City University of Hong Kong\\
  Kowloon, Hong Kong SAR, China\\
  \texttt{yannisun@cityu.edu.hk} \\
}

\begin{document}

\maketitle
\begin{abstract}
Prokaryotic viruses, which infect bacteria and archaea, are key players in microbial communities. Predicting the hosts of prokaryotic viruses helps decipher the dynamic relationship between microbes. Experimental methods for host prediction cannot keep pace with the fast accumulation of sequenced phages.
Thus, there is a need for computational host prediction. Despite some promising results, computational host prediction remains a challenge because of the limited known interactions and the sheer amount of sequenced phages by high-throughput sequencing technologies. The state-of-the-art methods can only achieve 43\% accuracy at the species level. In this work, we formulate host prediction as link prediction in a knowledge graph that integrates multiple protein and DNA-based sequence features.  Our implementation named CHERRY can be applied to predict hosts for newly discovered viruses and to identify viruses infecting targeted bacteria. We demonstrated the utility of CHERRY for both applications and compared its performance with 11 popular host prediction methods. To our best knowledge, CHERRY has the highest accuracy in identifying virus-prokaryote interactions. It outperforms all the existing methods at the species level with an accuracy increase of 37\%. In addition, CHERRY's performance on short contigs is more stable than other tools.
\textbf{Contact} yannisun@cityu.edu.hk\\
\end{abstract}


\newpage

\section{Introduction}
\label{sec:intro}
Prokaryotic viruses (shortened as viruses hereafter), including bacteriophages and archaeal viruses, are highly ubiquitous and abundant. They are key players in a wide range of microbial communities. By infecting their hosts, they can regulate both the composition and function of the microbiome. Thus, identifying the hosts of novel viruses play an essential role in characterizing the interactions of the organisms inhabiting the same niche. 

Despite the importance of understanding the virus-host interactions, the characterized interactions between viruses and prokaryotes is just the tip of the iceberg. Experimental methods for host identification cannot keep pace with the rapid accumulation of sequenced phages.
In addition, many experimental methods require cultivation of the prokaryotes\cite{dvzunkova2019defining}, limiting their large-scale applications. 
As reported in \cite{edwards2005viral, wawrzynczak2007global}, no more than 1\% of microbial hosts can be cultivated successfully in laboratories. With advancements of high-throughput sequencing, a large number of new phages can be sequenced from host-associated or environmental samples.  As a result, host identification for newly sequenced phages lagged behind the fast accumulation of the sequencing data. Thus, computational approaches for host prediction are in great demand. 

There are two major challenges for computational host prediction. The first one is the lack of known virus-host interactions. For example, the number of known interactions dated up to 2020 only accounted for \textasciitilde 40\% (1,940) of the prokaryotic viruses at the NCBI RefSeq at that time. Meanwhile, among the 60,105 prokaryotic genomes at the NCBI RefSeq, only 223 of them have annotated interactions with the 1,940 viruses.
The limited known interactions require carefully designed models or algorithms for host prediction. Second, although sequence similarity between viruses and hosts is an insightful feature for host prediction, not all viruses share common regions with their host genomes. For example, in the RefSeq database, \textasciitilde 24\% viruses do not have significant alignments with their hosts. Therefore, the alignment-based methods cannot identify hosts for some phages. 

\begin{table*}[]
\begin{tabular}{lp{3cm}p{7.5cm}cp{2.5cm}}
\hline
Name      & \multicolumn{1}{c}{Model}                           & \multicolumn{1}{c}{Description}                                                                                                                                                                                                                                                                                                                   & Prediction level \\ \hline
WIsH \cite{galiez2017wish}      &  Markov model                        & Trained a homogeneous Markov model for   potential host genomes and calculated the likelihood of a prokaryote genome as   the host for a query virus.                                                                                                                                                                             & Genus            \\
PHP \cite{lu2021prokaryotic}       & Gaussian mixture model                   & Utilized the k-mer frequency, which can reflect the codon usage patterns shared by the viruses and the hosts, to train a Gaussian mixture model.                                                                                                                                                                          & Genus            \\
HoPhage \cite{tan2021hophage}  & Deep learning model and Markov chain   algorithm & Used the CDS of each candidate host   genome to construct a Markov chain model and then calculated the likelihood of   query phage fragments infecting the candidate host genomes. They also use a   deep learning model and finally integrated the results of deep learning model   with the Markov model for host prediction. & Genus            \\
VPF-Class \cite{pons2021vpf} & Sequence match-based model                      & Utilized Viral Protein Families (VPFs)   downloaded from the IMG/VR system to estimate the similarity between query   viruses and viruses with known hosts.                                                                                                                                                                    & Genus            \\
RaFAH \cite{coutinho2021rafah}     & Random forest model                        & Used MMseqs2 to generate protein   clusters and constructed profile hidden Markov models (HMMs). Then, they used   features output by the HMM alignments and trained a multi-class random forest   model.                                                                                                                           & Genus            \\
HostG \cite{shang2021detecting}    & Graph convolutional network (GCN)          & Utilized the shared protein clusters   between viruses and prokaryotes to create a knowledge graph and trained a GCN for prediction.                                                                                                                                                                                           & Genus            \\
PHIST \cite{PHIST}    & Alignment-based model                      & Predicted prokaryotic hosts of   viruses based on exact matches between viral and host genomes.                                                                                                                                                                                                                                 & Species          \\
PredPHI \cite{li2020deep} & Convolutional neural network (CNN) & Utilized chemical component information, such as abundance of the amino acids, from protein sequences to train a CNN for host prediction. & Species\\ 
PHIAF \cite{PHIAF}    & Generative adversarial network (GAN) and convolutional neural network (CNN)         & Used the features originated from DNA and protein sequences, such as k-mer frequency and molecular weight, to train a CNN for host prediction. They also applied GAN to generate pseudo virus-host pairs from known virus-host interactions to enlarge the dataset.                                                                                                                                 & Species          \\
vHULK \cite{amgarten2020vhulk}     & Multi-layer perceptron models (MLP)        & Formulated host prediction as a multi-class classification problem where the inputs are viruses and the   labels are the prokaryotes. The features used in their model are the protein   profile alignment results against pVOGs database of phage protein families.                                                         & Species          \\
DeepHost \cite{DeepHost} & Convolutional neural network (CNN)         & Designed a genome encoding method   to encode genomes of various lengths into 3D matrices using k-mer features and trained a CNN model   for host prediction.                                                                                                                                                                    & Species          \\
VHM-net \cite{wang2020network}  & Markov random field                        & Utilized the Markov random field   framework for predicting whether a virus infects a target prokaryote by   combining multiple features between viruses and prokaryotes such as CRISPRs,   the output score of WIsH, BLASTN alignments, etc.                                                                                  & Species          \\ 

CHERRY & Graph convolutional encoder and decoder for link prediction & Formulated the host prediction problem as a link prediction problem in a multimodal graph and designed an encoder-decoder structure for host prediction. The multimodal graph integrates of different types of features, including protein organization, CRISPR, sequence similarity, and k-mer frequency, into the nodes and edges. The edges connect viruses and prokaryotes from labeled (training) and unlabeled (test) data.  & Species

\\\hline

\end{tabular}
\caption{A comparison of models/methods employed for host prediction of prokaryotic viruses.}
\label{tab:compare}
\end{table*}

\subsection{Related work}
\label{sec:relate}
Multiple attempts have been made to predict hosts for viruses \cite{roux2021global, edwards2016computational, galiez2017wish, ahlgren2017alignment, wang2020network}. According to the design, these methods can be roughly divided into two groups: alignment-based and learning-based.  Alignment-based methods utilize the similarities between viruses or the similarities between viruses and prokaryotic genomes for host identification.  For example, VPF-Class \cite{pons2021vpf} utilizes Viral Protein Families (VPFs) downloaded from the IMG/VR system to estimate the similarity between query viruses and viruses with known hosts. According to the alignment results with VPFs, VPF-class can return predictions for each query contig. PHIST \cite{PHIST} utilizes sequence matches between viral and prokaryotic genomes for host prediction. By identifying the common $k$-mers shared by viral and prokaryotic genomes, PHIST estimates the probability of a virus-prokaryote pair forming a real interaction.

Alignment-based tools also use CRISPR for host prediction. Some prokaryotes keep a record of virus infection via CRISPR \cite{edwards2016computational}to prevent recurring infections \cite{achigar2017phage}. Thus, local alignment programs such as BLAST \cite{johnson2008ncbi} can be employed to predict hosts by searching for short matches between prokaryotes and viruses. However, it is estimated that only \textasciitilde 20\% of sequenced prokaryotes contain CRISPRs \cite{burstein2016major, grissa2007crisprdb}. Although CRISPR is an informative host prediction signal, many viruses do not have alignment results with the annotated or predicted CRISPRs of prokaryotes and thus cannot use this signal for host prediction.
 
  
Learning-based methods are more flexible. Most of these methods learn sequence-based features for host prediction. For example, WIsH \cite{galiez2017wish} trains a homogeneous Markov model for potential host genomes. The model then calculates the likelihood of a prokaryote genome as the host for a query virus and assigns the host with the highest likelihood. vHULK \cite{amgarten2020vhulk} formulates host prediction as a multi-class classification problem where the inputs are viruses and the labels are the prokaryotes. The features used in their deep learning model is the protein profile alignment results against pVOGs database of phage protein families \cite{grazziotin2016prokaryotic}. 
Rather than using the public database, RaFAH \cite{coutinho2021rafah} uses MMseqs2 \cite{steinegger2017mmseqs2} to generate protein clusters and constructs profile hidden Markov models (HMMs). Then, they use features output by the HMM alignments and train a multi-class random forest model. HoPhage \cite{tan2021hophage}, another multi-class classification model-based host prediction tool, uses deep learning and Markov chain algorithm. They use the CDS of each candidate host genome to construct a Markov chain model and then calculate the likelihood of query phage fragments infecting the candidate host genomes. They also use a deep learning model and finally integrate the results of deep learning model with the Markov model for host prediction. On the other hand, PHP \cite{lu2021prokaryotic} utilizes the $k$-mer frequency, which can reflect the codon usage patterns shared by the viruses and the hosts \cite{gouy1982codon, carbone2008codon}. DeepHost \cite{DeepHost} and PHIAF \cite{PHIAF} also utilize $k$-mer-based features to train a convolutional neural network for host prediction. 
Boeckaerts et al. build learning models using features extracted from receptor-binding proteins (RBPs) for host prediction \cite{boeckaerts2021predicting}. However, it is not trivial to annotate RBPs in all viruses. The authors only collected RBPs related to 9 hosts and thus this tool can only predict very limited host species.
HostG \cite{shang2021detecting} utilizes the shared protein clusters between viruses and prokaryotes to create a knowledge graph and trains a graph convolutional network for prediction. Although it has high accuracy of prediction, it can only predict the host at the genus level. 
The best host prediction performance at the species level is reported by VHM-Net \cite{wang2020network}, which incorporates multiple features between viruses and prokaryotes such as CRISPRs, the output score of WIsH, BLASTN alignments, etc. By combining these features, VHM-net utilizes the Markov random field framework for predicting whether a virus infects a target prokaryote. Nevertheless, the accuracy at the species level is only 43\%. 

\begin{figure*}[h!]
    \centering
    \includegraphics[width=0.8\linewidth]{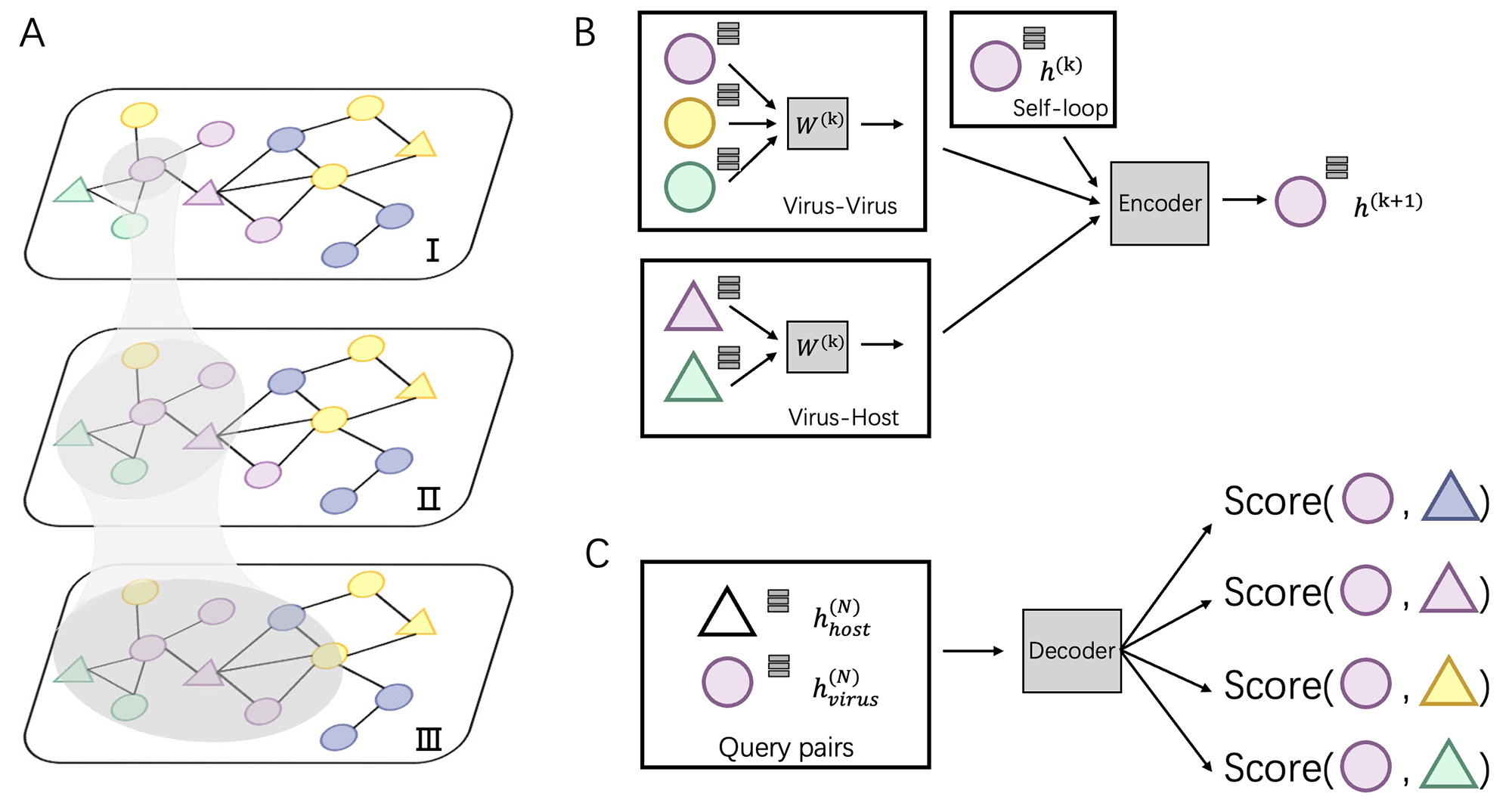}
    \caption{The key components of CHERRY. A) The multimodal knowledge graph. Triangle represents the prokaryotic node and circle represents virus nodes. Different colors represents different taxonomic labels of the prokaryotes. \uppercase\expandafter{\romannumeral1}-\uppercase\expandafter{\romannumeral3} illustrate graph convolution using neighbors of increasing orders. B) The graph convolutional encoder of CHERRY. C) The decoder of CHERRY.}
    \label{fig:model}
\end{figure*}

\subsection{Overview}
In this work, we develop a new method, CHERRY, which can predict the hosts' taxa (phylum to species) for newly identified viruses. First, we construct a multimodal graph that incorporates multiple types of interactions, including protein organization information between viruses, the sequence similarity between viruses and prokaryotes, and the CRISPR signals (Fig. \ref{fig:model} A). In addition, we use $k$-mer frequency as the node features to enhance the learning ability. Second, rather than directly using these features for prediction, we design an encoder-decoder structure to learn the best embedding for input sequences and predict the interactions between viruses and prokaryotes. The graph convolutional encoder (Fig. \ref{fig:model} B)  utilizes the topological structure of the multimodal graph and thus, features from both training and testing sequences can be incorporated to embed new node features. Then a link prediction decoder (Fig. \ref{fig:model} C) is adopted to estimate how likely a given virus-prokaryote pair forms a real infection. Unlike many existing tools, CHERRY can be flexibly used in two scenarios. It can take either query viruses or prokaryotes as input. For viruses, it can predict their hosts. For input prokaryotes, 
it can predict the viruses infecting them. Another feature behind the high accuracy of CHERRY is the construction of the negative training set. The dataset for training is highly imbalanced, with the real host as the positive data and all other prokaryotes as negative data. We carefully addressed this issue using negative sampling \cite{mikolov2013distributed}. Instead of using a random subset of the negative set for training the model, we apply end-to-end optimization and negative sampling to automatically learn the hard cases during training. To demonstrate the reliability of our method, we rigorously tested CHERRY on multiple datasets including the RefSeq dataset, simulated short contigs, metagenomic datasets. We compared CHERRY with WIsH, PHP, HoPhage, VPF-Class, RaFAH, HostG, vHULK, PHIST, DeepHost, PHIAF, and VHM-net, whose brief descriptions can be found in Table \ref{tab:compare}. The results show that CHERRY competes favorably against the state-of-the-art tools and yields 37\% improvements at the species level.

\section{Method}
We formulate host prediction as a link prediction problem \cite{al2006link} on a multimodal graph, which encodes virus-virus and virus-prokaryote relationships. To be specific, these relationships can be represented by a knowledge graph $G = (V, E)$ with node $v_i \in V$, where $i=1,2,..., N$. An edge between $v_i$ and $v_j$ is denoted as a tuple $(v_i, v_j) \in E$. There are two kinds of nodes in the graph: viral nodes $p_i \in P$ and prokaryotic nodes $h_i \in H$ ($P \cup H = V$). Then the link prediction task can be defined as: \textbf{given} a viral node $p_i$ and a prokaryotic node $h_i$, \textbf{what is the probability} of $p_i$ and $h_i$ having a link (infection). In the following section, first, we will describe how we construct the multimodal graph. Then, we will introduce the encoder-decode structure of CHERRY.

\subsection{Construction of the knowledge graph $G$} 
\label{sec:graph}
To utilize the features from both training and testing samples, we construct a multimodal graph $G$ by connecting viruses and prokaryotes in both the reference database and the test set. This multimodal graph is composed of protein organizations, sequence similarity, and CRISPR-based similarity. The node in the graph encodes $k$-mer frequency feature from the DNA sequences. According to the type of the connections, the edges in the knowledge graph can be divided into virus-virus connections and virus-prokaryote connections.

\paragraph{Virus-virus connections:}
we utilize the protein organizations to measure the similarity of biological functions between viruses. Intuitively, if two viruses share similar protein organizations, they are more likely to infect the same host. First, we construct protein clusters using the Markov clustering algorithm (MCL) on all viral proteins. For reference viral genomes, the proteins are downloaded from NCBI RefSeq. For query contigs, we use Prodigal \cite{hyatt2010prodigal} to conduct gene finding and protein translation. Then, we employ MCL to cluster proteins with $inflation =2.0$ based on the DIAMOND BLASTP \cite{buchfink2015fast} comparisons (E-value \textless 1e-5). Second, we followed \cite{bolduc2017vcontact, shang2021bacteriophage} and use Eq. \ref{edge1} to estimate the probability of two viruses $X$ and $Y$ sharing at least $c$ common protein clusters by assuming that each protein cluster has the same chance to be chosen. $x$ and $y$ are the numbers of proteins in $X$ and $Y$, respectively. Because Eq. \ref{edge1} computes the background probability under the hypothesis that virus $X$ and $Y$ don't share common host, we will reject this hypothesis when P is smaller than a cutoff. Finally, only pairs with P( $\ge$ c) smaller than $\tau_1$ will form virus-virus connections (Eq. \ref{edge2}).

\begin{equation}
    \label{edge1}
    \indent P( \ge c) =  {\textstyle \sum_{i=c}^{min(x,y)}} \frac{\binom{x}{i}\binom{n-x}{y-i}}{\binom{n}{y}}
\end{equation}

\begin{equation}
\label{edge2}
\indent virus\raisebox{0mm}{-}virus = \begin{cases}
  1 & \text{ if  } P( \ge c)  < \tau_1  \\
  0 & \text{ otherwise } 
\end{cases}
\end{equation}

\paragraph{Virus-prokaryote connections:}
we apply the sequence similarity between viral and prokaryotic sequences to define the virus-prokaryote connections. There are two kinds of sequence similarity that can be employed: CRISPR-based and general local similarity. Some prokaryotes will integrate some viral DNA fragments into their own genomes to form spacers \cite{dutilh2014highly, roux2016ecogenomics} in CRISPR. Therefore, many existing tools have used CRISPR as a main feature for host prediction \cite{edwards2016computational, Nathan2020vhmnet}. In our method, CRISPR Recognition Tool \cite{bland8p} is applied to capture potential CRISPRs from prokaryotes. If a viral sequence shares a similar region with the CRISPR, we will connect this viral node to the prokaryotic node.

However, only a limited number of CRISPRs can be found. We thus also use BLASTN to measure the sequence similarity between the sequences. Viruses can mobilize host genetic material and incorporate it into their own genomes. Occasionally, these genes can bring an evolutionary advantage and the viruses will preserve them \cite{edwards2016computational}. For all viruses $p_i \in P$ and prokaryotes $h_i \in H$, we will run BLASTN for each pair $(p_i, h_i)$. Only pairs with BLASTN E-value smaller than $\tau_2$ will form virus-prokaryote connections. In addition, because we have known virus-prokaryote connections from the public dataset, we connect the viruses with their known hosts regardless of their alignment E-values. Finally, the edges $(p_i, h_i) \in E$ can be formulated as Eq. \ref{edge3}. If there is an overlap between CRISPR-based and BLASTN-based edge, we will only create one edge between the virus and prokaryote.

\begin{equation}
\label{edge3}
\indent virus\raisebox{0mm}{-}prokaryote = \begin{cases}
    & \text{ if  $\exists$ CRISPR alignment } \\
  1 & \text{ or  BLASTN }E_{value} <  \tau_2 \\
    & \text{ or  $\exists$ interaction in dataset}  \\
  0 & \text{ otherwise } 
\end{cases}
\end{equation}
\noindent Both $\tau_1$ and $\tau_2$ are the default parameters given in \cite{shang2021bacteriophage} and \cite{camacho2009blast+}.

\paragraph{Nodes construction:} 
The nodes in the multimodal graph represent the features from both viral and host genomes. In our work, we utilize the $k$-mer frequency as our node feature. Because the dimension of the feature vector increases exponentially with $k$,  we choose $k=4$ and obtain a 256-dimensional vector for each viral or host genome. 

\subsection{The encoder-decoder framework in CHERRY}
After constructing the multimodal graph $G$, we will feed it to our model for training and prediction. Given a virus-prokaryote pair $(p_i, h_i)$, our aim is to determine how likely there is a link (infection) between virus $p_i$ and prokaryote $h_i$. Towards this goal, we develop a non-linear, multi-layer decoder-encoder network CHERRY that operates on the multimodal graph $G$. CHERRY has two main components:

\begin{itemize}
\item[$\bullet$] \textit{Graph convolutional encoder:} a graph convolutional network operating on $G$ and embedded node features in $G$ (Fig. \ref{fig:model} B)
\item[$\bullet$] \textit{Query pairs decoder:} a 2-layer neural network classifier that can output a probability score for each virus-prokaryote pair (Fig. \ref{fig:model} C)
\end{itemize}

\noindent Detailed information about the two components and the end-to-end training method will be discussed in the following sections.

\subsubsection{Graph convolutional encoder}
The input of the encoder is the multimodal graph $G$. Because we have both labelled (viruses with known hosts) and unlabelled (viruses without known hosts) nodes in the graph, the main idea of the encoder is to utilize the topological structure to propagate information from labelled nodes to unlabelled nodes. The output of the encoder will be $d$-dimensional embedded vectors that integrate virus-virus similarity and virus-prokaryote similarity gained from the multimodal graph $G$.

We will use graph convolutional neural network (GCN) to conduct feature embedding of the nodes. GCN is one well-studied model for graph data. Recently, it has some successful applications in biological data \cite{shang2021bacteriophage, cramer2021alphafold2}. In our encoder, we will take advantage of the feature embedding of GCN to encode the feature vectors for viruses and prokaryotes. Specifically, for a given node (Fig. \ref{fig:model} A\uppercase\expandafter{\romannumeral1}), our encoder performs convolution on its neighbors' node vectors and itself. In each convolution operation, the encoder considers the 1-step neighborhood of the nodes (Fig. \ref{fig:model} A\uppercase\expandafter{\romannumeral2}) and applies the same transformation to all nodes in the graph. Then the successive convolution will be applied in the $l$ layers, and finally, each node will effectively convolve the information from its $l$-step neighborhood (Fig. \ref{fig:model} A\uppercase\expandafter{\romannumeral3}). $l$ is the number of the graph convolutional layers in the encoder. A single graph convolutional layer can be represented as Eq. \ref{gcn1}.

\begin{equation}
    \label{gcn1}
    \indent h^{l+1} = \phi (\tilde{D}^{-\frac{1}{2}} \tilde{A} \tilde{D}^{\frac{1}{2}}h^{(l)}W^{(l)})
\end{equation}

\noindent $\phi$ is the activation function in the graph convolutional layer. $A \in \mathbb{R}^{N \times N}$ is the adjacency matrix, where $N$ is the number of nodes in the multimodal graph. $\tilde{A} = A + I$, where $I \in \mathbb{R}^{N \times N}$ is the identity matrix. $\tilde{D}$ is the diagonal matrix calculated by $D_{ii} = {\textstyle \sum_{j}} \tilde{A}_{ij}$. $h^{(l)}$ is the hidden feature in the $l$th layer and $h^{(0)} \in \mathbb{R}^{N \times 256}$ is the 4-mer frequency vector of each node. $W^{l}$ is a matrix of the trainable filter parameters in the $l$-layer. Finally, the encoder will output a $d$-dimensional embedded vector, which encoded prior knowledge from $l$-step neighborhood for each node in the graph. $d$ is the dimension of $W$ in the output layer. Because the convolutional layer will be conducted on all nodes in $G$, features from both training and testing samples can be used in the encoder to enhance the learning ability.

\subsubsection{Decoder for link prediction}
After encoding the feature vectors for viral and host genomes, we apply a 2-layer neural network classifier to decode the embedded vectors outputted from the encoder. This decoder aims to judge how likely these query pairs form actual infections. Thus, the input of the decoder is a query set $Q$, and the output of the decoder is a probability score. Each element in $Q$ is called a query vector $q_{ij}$ and is calculated by Eq. \ref{decode1}.

\begin{equation}
    \label{decode1}
    \indent  q_{ij} = encoder(p_i) - encoder(h_j)
\end{equation}

First, we generate all-against-all virus-prokaryote pairs and calculate all query vectors $q_{ij} \in Q$. $encoder(\cdot)$ represents the output of graph convolutional encoder. $p_i \in P$ represents the virus node, and $h_j \in H$ represents the prokaryotic node. Then we employ a 2-layer neural network to decode the feature vector for each input $q_{ij}$ as shown in Eq. \ref{decode2}. 

\begin{equation}
    \label{decode2}
\indent      \left\{\begin{matrix}
 q_{ij}^{(l+1)} = \phi (q_{ij}^{(l)} \theta^{(l)})\\
 decoder(q_{ij}) = sigmoid(q_{ij}^{(L-1)})
\end{matrix}\right.
\end{equation}

\noindent $q_{ij}^{(l)}$ is the hidden feature in the $l$th layer. $q_{ij}^{(0)} = q_{ij}$ and $L$ is the maximum number of the layers in the network. $\phi$ is the activation function. $decoder(\cdot)$ represents the output of the link prediction decoder. Because the activation function of the output layer is the $sigmoid$ function, $decoder(\cdot)$ can be used as the probability score for each pair. 

\subsubsection{Model training} 
Recent results show that the modeling of topological structure data can be greatly improved through end-to-end learning \cite{defferrard2016convolutional,gilmer2017neural}. Thus, we optimize overall trainable parameters for CHERRY and backpropagate loss on the multimodal graph. The trainable parameters of CHERRY are: (i) convolutional filters' weight matrices $W$ in the encoder and (ii) query parameter matrices $\theta$ in the decoder.

\begin{equation}
    \label{opt}
    \indent J(i, j) = -logP_{r}^{ij} - \mathbb{E}_{n\sim P_{r}}log(1 - P_{r}^{ik})
\end{equation}
  
There are two kinds of query pairs that will be generated by Eq. \ref{decode1}: positive pairs and negative pairs. Positive pairs represent known virus-prokaryote interactions given by the dataset. Negative pairs represent the pairs with no evidence for interaction. Then, during the training process, we optimize the model using the cross-entropy loss as shown in Eq. \ref{opt}. This equation encourages the model to assign higher probabilities to positive pairs $(p_i, h_j)$ than the randomly created negative pair $(p_i, h_k)$. 

Because we form all-against-all query pairs from all viruses and hosts, the number of negative query pairs will be much larger than the positive query pairs. To solve this problem, rather than sampling a subset of the negative pairs, we optimize the model through negative sampling, a method introduced in recent publications \cite{mikolov2013distributed, trouillon2016complex}. Negative sampling allows us to train the model using hard cases. When calculating the negative sampling loss in training, we replace the prokaryotic node $h_j$ in a positive pair $(p_i,h_j)$ with another prokaryotic node $h_k$ (thus forming a negative sample) according to a sampling distribution $P_{r}$ defined in \cite{mikolov2013distributed}. 
Intuitively, if a negative sample has a higher loss, it will have a higher probability to be selected for training. Thus, the negative sampling method helps enhance the learning ability of the model compared to sub-sampling because the latter cannot represent the real distribution of the sample space, especially when the number of negative samples is much larger than the number of positive samples. With negative sampling, CHERRY can automatically learn a more accurate boundary using the hard cases.

To ensure that the optimization process can learn as many query pairs as possible during the negative sampling process, we train it for a maximum of 250 epochs using the Adam optimizer \cite{kingma2014adam} with a 0.01 learning rate to update the parameters. Before evaluating our model, we fixed the random seed in the program to ensure that we had the same initial parameters. To guarantee the model's reliability, we also save the parameters so that users can directly load the parameters when applying CHERRY on their own dataset. Users can also use the parameters as a pre-trained model and add more interactions for training, which will help the model converge faster. 

\subsection{Experimental setup}
\label{sec:setup}
This section will introduce how we evaluate our model and compare it with the state-of-the-art tools. 

\subsubsection{Metrics}
According to the usage of link prediction, CHERRY can be employed in two different scenarios: 1) predicting host for newly identified viruses; and 2) identifying viruses that infect targeted 
prokaryotes. Thus, we apply two different evaluation methods, respectively.

\paragraph{Scenario 1: predicting hosts for newly identified viruses} We use the same experimental setup and metrics as the previous works to ensure consistency and a fair comparison. Following the previous work, one virus is assumed to infect only one species, which is not always true but is a commonly used assumption for evaluation. Thus, in the experiments, for each test virus $p_i$, we will compute the score between $p_i$ and all prokaryotes in set $H$ and output the prokaryote with the highest decoder score (Eq. \ref{m1}).

\begin{equation}
    \label{m1}
    \indent \mathop{\arg\max}_{h_j \in H}\ decoder(q_{ij}) 
\end{equation}

\noindent 
$q_{ij}$ is defined in Eq. \ref{decode1}. We predict hosts for all viruses in the test set and evaluate the performance by checking whether the predicted prokaryotes are from the same taxon (such as same species) as the known interactions. The ratio of correct prediction is the \textit{accuracy}, which has the same definition as previous works. In addition, because the graph contains all the potential hosts (prokaryotes) and the number of predictions is equal to the number of test viruses, the \textit{recall} and \textit{precision} are the same as the \textit{accuracy}. 

\paragraph{Scenario 2: identifying viruses that infect targeted prokaryotes} The goal is to identify which viruses can infect a specified prokaryote. 
Because one prokaryote can be infected by multiple viruses, we set a threshold $\mu$ to output a set of candidate viruses. 

\begin{equation}
    \label{m2}
    \indent S(h_j)= \left \{  p_i \in P \mid \ decoder(q_{ij}) > \mu  \right \}
\end{equation}

\noindent As shown in Eq. \ref{m2}, for each specified prokaryote, we will predict a set of viruses whose interaction probabilities (calculated by the decoder) are larger than $\mu$. Unlike the host prediction task, this equation might return viruses infecting different prokaryotes as long as the probability is larger than the threshold. Then, for each prokaryote $h_j$, the \textit{precision} represents how many viruses in $S(h_j)$ truly infect $h_j$ based on the ground truth. The \textit{recall} represents how many viruses infecting $h_j$ are included in $S(h_j)$.

\subsubsection{Dataset}
\begin{figure}[h!]
    \centering
    \includegraphics[width=0.55\linewidth]{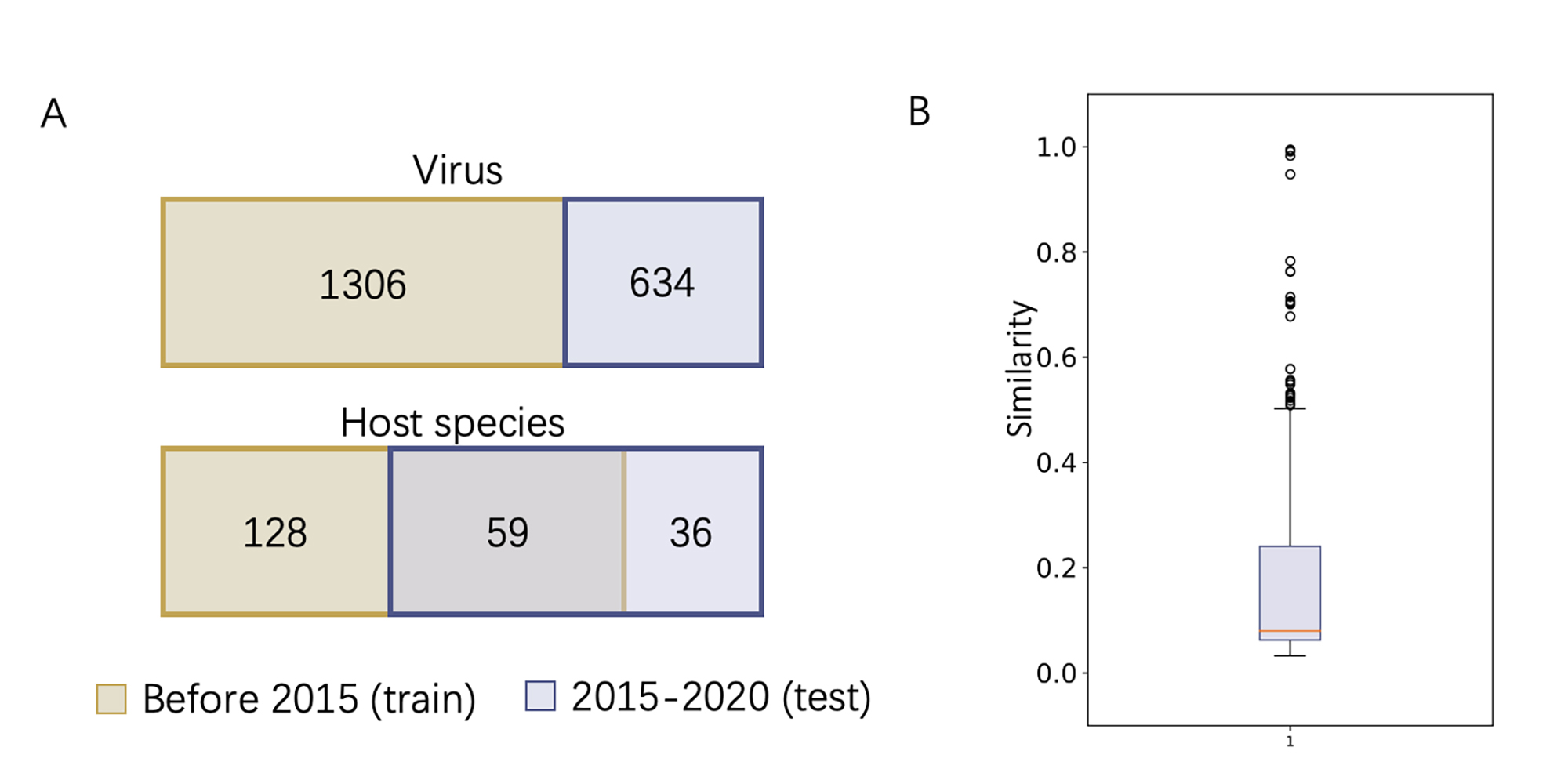}
    \caption{The benchmark dataset for virus-host interactions. (A) The viruses and their hosts in the training and test set, respectively.  (B) Similarity between viruses in the training and test sets.}
    \label{fig:data}
\end{figure}

we followed \cite{lu2021prokaryotic} and used the same virus-host relationship benchmark dataset for training (the VHM dataset) and testing (the TEST dataset). This benchmark dataset contains viruses and prokaryotic genomes that were deposited to NCBI RefSeq in or before 2020. The detailed information is shown in Fig. \ref{fig:data} (A). Following \cite{lu2021prokaryotic}, we download all 1,940 viruses from the NCBI RefSeq database and separate the training set and test set according to their submission time (before and after 2015). Thus, we have 1,306 positive pairs for training and 634 positive pairs for testing, respectively. Although every virus is unique, some of them infect the same host. The training and test set share 59 host species. To show the overall similarity between the training and test set, we use Dashing \cite{baker2019dashing} to estimate the sequence similarity between viruses. We record the largest value for each test virus against all training viruses and report the overall similarity in Fig. \ref{fig:data} (B). The result reveals that only a few test viruses are similar to the training viruses, with the mean similarity being 0.1. Along with 233 host species, we also have 60,105 prokaryotic genomes obtained from the NCBI genome database. 

To further assess the utility of CHERRY on predicting hosts for newly discovered viruses, we applied it to viruses from three recently published metagenomic datasets. 
The information of the datasets are summarized as below:

\begin{itemize}
\item[$\bullet$] \textit{MetaHiC data from healthy human gut}: 6545 host-phage interactions were identified from human gut metagenomic samples using a proximity ligation-based approach named MetaHiC \cite{marbouty2021metahic}. The dataset is available at: \url{https://github.com/mmarbout/HGP-Hi-C}.
\item[$\bullet$] \textit{Glacier metagenomic dataset}: a newly published metagenomic dataset sampled from the ice core on the Tibetan Plateau \cite{zhong2021glacier}. The authors present 33 new phage contigs and four dominant bacteria genus isolated from the sample. The dataset is available at: \url{https://datacommons.cyverse.org/browse/iplant/home/shared/iVirus/Tibet_Glacier_viromes_2017}.
\item[$\bullet$] \textit{Gut metagenomic dataset}: 3,738 previously unknown phages discovered from human gut metagenomic data \cite{benler2021thousands}. These phage genomes represent 451 putative genera whose hosts remain unknown. The dataset is available at: \url{ftp://ftp.ncbi.nih.gov/pub/yutinn/benler_2020/gut_phages/}.
\end{itemize}

\section{Result}
\label{sec:exp}
In this section, we will show our experimental results on multiple datasets and compare CHERRY against 11 host prediction tools: WIsH \cite{galiez2017wish}, PHP \cite{lu2021prokaryotic}, Hophage \cite{tan2021hophage}, VPF-Class \cite{pons2021vpf}, HostG \cite{shang2021detecting}, RaFAH \cite{coutinho2021rafah}, PHIST \cite{PHIST}, PHIAF \cite{PHIAF}, vHULK \cite{amgarten2020vhulk}, DeepHost \cite{DeepHost}, and VHM-Net \cite{wang2020network}. We also recorded the host prediction performance using either BLASTN or CRISPRs, which are two frequently used features for identifying the interactions. The experimental results consist of two parts.

\paragraph{Usage 1: predicting hosts for viruses}
First, we will report the host prediction performance across different taxonomic levels (from species to phylum) on the benchmark dataset. We will analyze how the knowledge graph and the encoder-decoder structure impact the host prediction accuracy. In addition, we will investigate why the alignment-based method fails to perform well in the host prediction task. Second, because many metagenomic assemblies contain short contigs, we will evaluate the performance of CHERRY with short contigs as input. Third, we will visualize the results of host prediction on different viral families and analyze data-related challenges for species-level host prediction. Then, we present how CHERRY can be used to narrow the search scope of the potential hosts for newly identified viruses. Finally, we will show the application of CHERRY on predicting hosts for newly discovered phage genomes/contigs from three metagenomic datasets.

\paragraph{Usage 2: predicting viruses that infect prokaryotes}
In this part, we will present the performance of CHERRY on identifying viruses that can infect given prokaryotes.

\begin{figure*}[h!]
    \centering
    \includegraphics[width=0.75\linewidth]{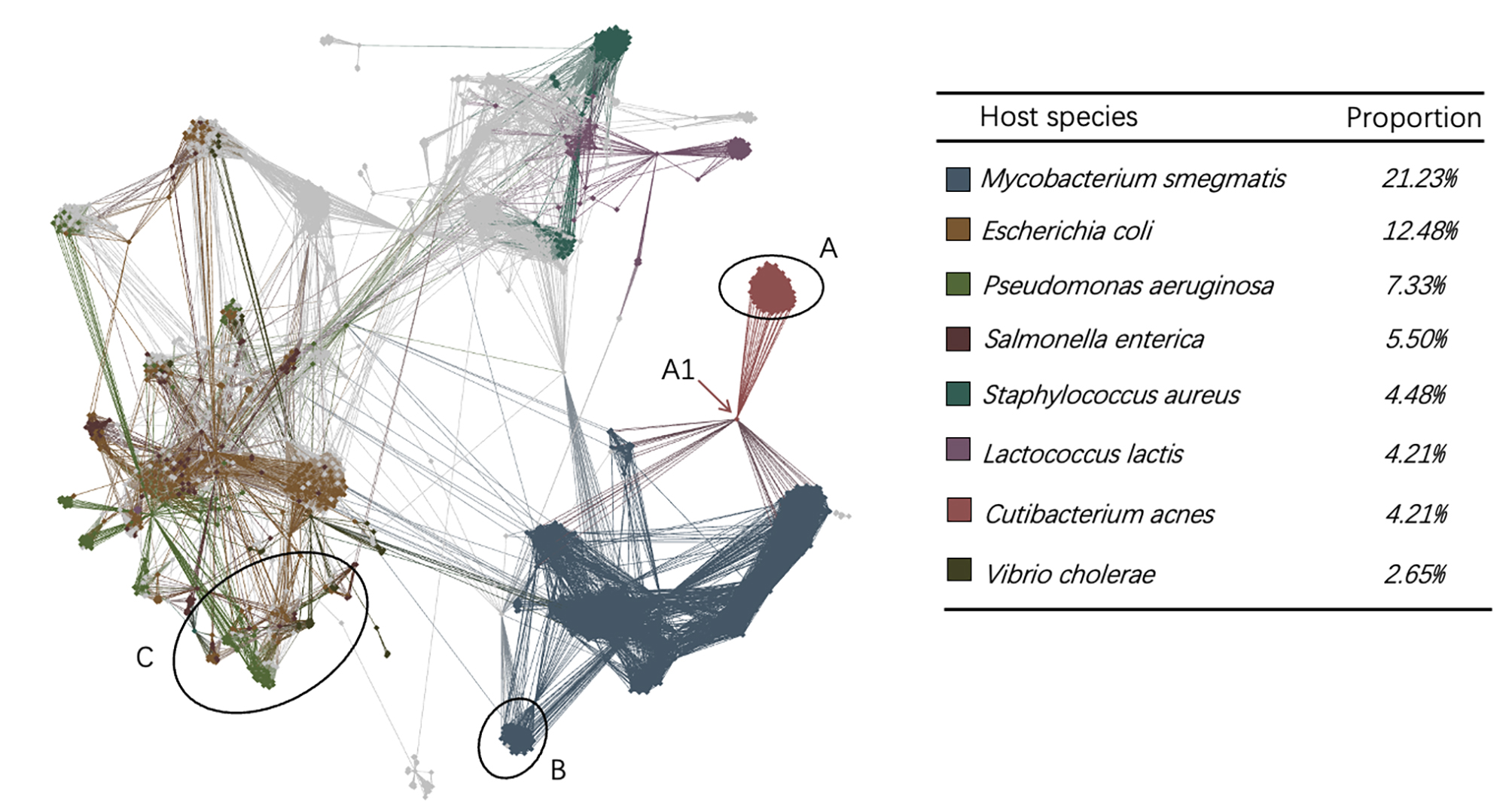}
    \caption{Visualization of the multimodal graph. The colors of the nodes represent their labels. For prokaryotic nodes, the labels represent their species. For virus nodes, the labels represent their hosts' species. Because there are a large number of labels, this graph only colors the top 8 labels with the largest number of nodes. All others are gray. }
    \label{fig:network}
\end{figure*}

\subsection{Visualization of the knowledge graph}
We use Gephi \cite{bastian2009gephi} to visualize the topological structure of the knowledge graph in Fig. \ref{fig:network}. 
The GCN encoder utilizes the local topology of the graph to embed the node features. If the edges only connect nodes with the same label, a simple label propagation method based on graph connectivity can achieve accurate predictions. However, although there are some clusters containing nodes with consistent labels (e.g. region A and region B in Fig. \ref{fig:network}), some subgraphs are mixed with multiple labels. For example, region C contains more than six labels. In addition, we also found that prokaryotic nodes usually connect to the virus nodes with different labels. For example, node A1 represents \textit{Cutibacterium acnes}; but it has many edges (alignments) with phages that infect \textit{Mycobacterium smegmatis}. In this case, simple alignment-based or label propagation methods will fail to make the correct host prediction. But the GCN-based encoder can integrate both the sequence similarity, $k$-mer composition, and the topological information for making correct predictions using this graph.

\subsection{Predicting hosts for prokaryotic viruses}
In this experiment, we used the benchmark dataset associated with Fig. \ref{fig:data} for training and evaluation. For each test virus, our predicted host is the one with the highest prediction score out of the 60,105 prokaryotes. Then we calculate the accuracy according to the known hosts in the benchmark dataset.

\begin{figure*}[h!]
    \centering
    \includegraphics[width=0.98\linewidth]{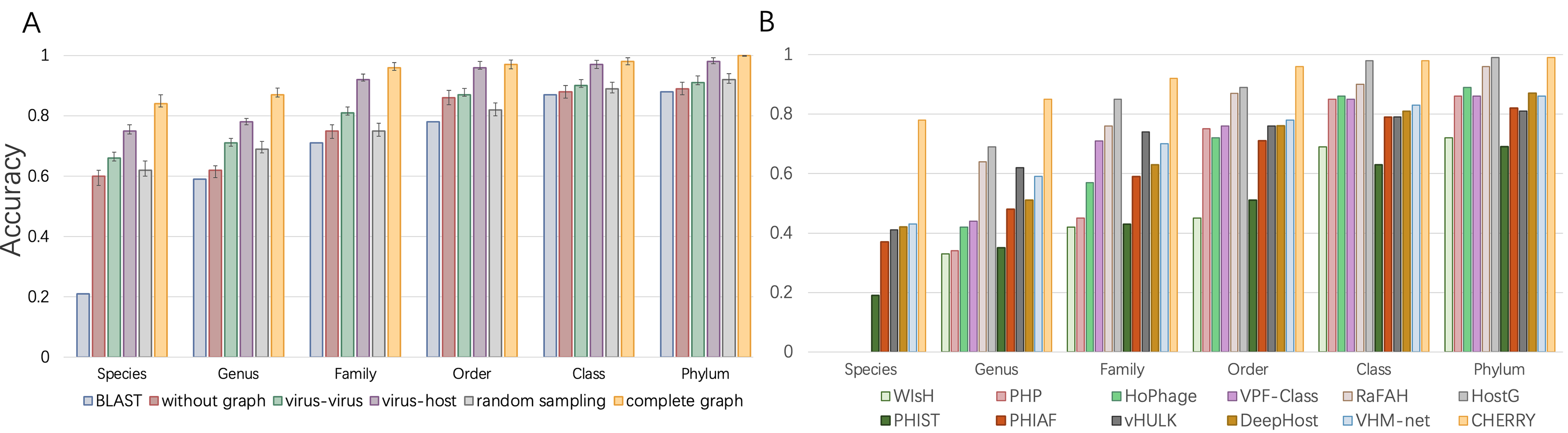}
    \caption{Host prediction performance on the benchmark dataset. Y-axis: accuracy. (A) 10-fold cross-validation on the training set. X-axis represents different types of graphs and BLAST-based host prediction. \textit{without graph}: training with only decoder.  \textit{Virus-virus}: the knowledge graph only contains \textit{virus-virus} edges. \textit{virus-prokaryote}: the knowledge graph only contains \textit{virus-prokaryote} edges. \textit{random sampling}: the model is trained on the complete graph with a randomly sampled negative set. \textit{complete graph}: the model is trained on the complete graph with negative sampling. Error bar represents the highest, lowest, and average accuracy of the 10-fold cross-validation. (B) Comparison of host prediction accuracy on the test set from species to phylum. Tools that can output predictions at the species level (PHIST, PHIAF, vHULK, DeepHost, VHM-net, and CHERRY) are grouped together and ordered based on their species-level performance.}
    \label{fig:train_valid}
\end{figure*}

\subsubsection{Cross-validation results and ablation study}
To train a reliable model and evaluate the overall learning ability of CHERRY, we applied 10-fold cross-validation on the training set. First, we randomly split the training set into ten subsets. Then, we trained the model on nine subsets, validated it on the tenth iteratively, and recorded the prediction performance. As shown in Fig. \ref{fig:train_valid}(A), we reported the highest, lowest, and average performance of the 10-fold cross-validation from the species to phylum level. We also reported the BLAST results of the host prediction. Specifically, we used 60,105 prokaryotes as reference genomes and then conducted alignments between test viruses and the reference set. Then we predict the host using k nearest neighbor (KNN) with $k = 1$ (the best alignment) or $k > 1$ (majority vote). Here, we only report the results based on the best alignment because it has higher accuracy. Fig. \ref{fig:train_valid}(A) shows that CHERRY largely improves the performance at the species level. Also, it is worth noting that BLASTN and CRISPRs can only return results for \textasciitilde 65.5\% and \textasciitilde 24.6\% viruses, respectively. However, CHERRY can predict hosts for all viruses. 

\paragraph{Ablation study} In addition, we investigated how different components in the knowledge graph affect the prediction accuracy. Our knowledge graph contains two types of edges: virus-virus and virus-prokaryote. We tested CHERRY by removing the graph, only keeping one type of edge in the graph, and using the complete graph. The comparison is shown in Fig. \ref{fig:train_valid}(A).

\begin{itemize}
\item[$\bullet$] \textit{without graph}: We trained the model without graph (without encoder). The decoder only uses the $k$-mer frequency vectors as inputs. 
\item[$\bullet$] \textit{virus-virus}: We trained the model with a graph that only contains virus-virus edges. 
\item[$\bullet$] \textit{virus-prokaryote}: We trained the model with a graph that only contains virus-prokaryote edges. 
\item[$\bullet$] \textit{complete graph}: We trained the model with the complete multimodal graph.
\end{itemize}

The results show that both virus-virus similarity and virus-prokaryote similarity can enhance the learning ability and improve the performance. The complete knowledge graph that contains two types of edges achieves the best performance. 
We also show the comparison between training with random sampling and negative sampling in Fig. \ref{fig:train_valid}(A). The prediction results show that negative sampling can largely improve the host prediction accuracy.

\subsubsection{Evaluation on the test set}
We fixed the model parameters based on the highest validation accuracy and applied the trained model to host prediction for test viruses. To conduct a fair comparison, we also re-trained the learning-based models on the same training set and applied them on the same test set. Because VPF Class is an alignment-based method, we used their database for host prediction. As shown in Fig. \ref{fig:train_valid} (B), CHERRY achieved the best accuracy of 78\% in predicting the hosts' species, which is 35\% higher than the next best tool VHM-net. To prove that the convolutional graph encoder can enhance the learning ability of CHERRY, we reused the model \textit{without graph} in Fig. \ref{fig:train_valid}(A) for host prediction. The final result on the test set decreases to 56\%, and thus, the convolutional graph encoder helps host prediction. 


\subsubsection{Impact of training vs test similarity on the prediction performance}
\begin{figure}[h!]
    \centering
    \includegraphics[width=0.55\linewidth]{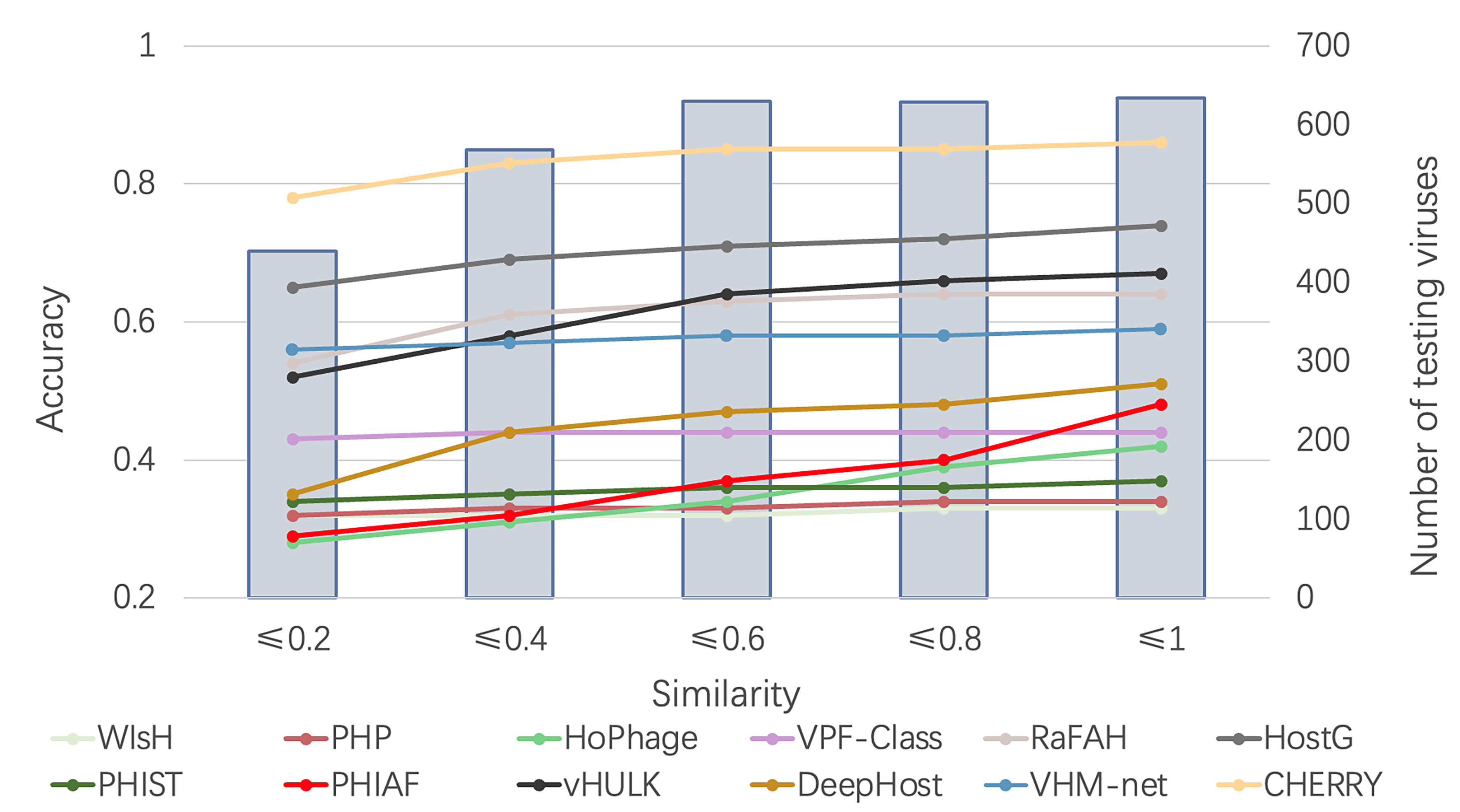}
    \caption{The impact of training-vs-test sequence similarity on host prediction at the genus level. X-axis: maximum dashing similarity. Left Y-axis: accuracy (line segments). Right Y-axis: number of test viruses under each similarity cutoff (gray bars). }
    \label{fig:similarity}
\end{figure}

In the \textit{Dataset} section, we used Dashing to measure the similarity between the test set and training set. To show how the similarity affects the host prediction performance, we divided the test set according to the dashing similarity. We recorded the accuracy at the genus level in Fig. \ref{fig:similarity}. X-axis stands for the maximum similarity between genomes in the training set and test set. For example, X-axis value 0.8 indicates that all the genomes in the test set have similarity $\leq$ 0.8 against the genomes in the training set. Fig. \ref{fig:similarity} also shows how the similarity influences other tools.

As expected, with the increase of the similarity, more test genomes with high similarities are included, and the accuracy of all methods increases. The gap between CHERRY and all other methods clearly shows that our model outperforms the state-of-the-art tools on a wide range of similarities.

\begin{figure}[h!]
    \centering
    \includegraphics[width=0.55\linewidth]{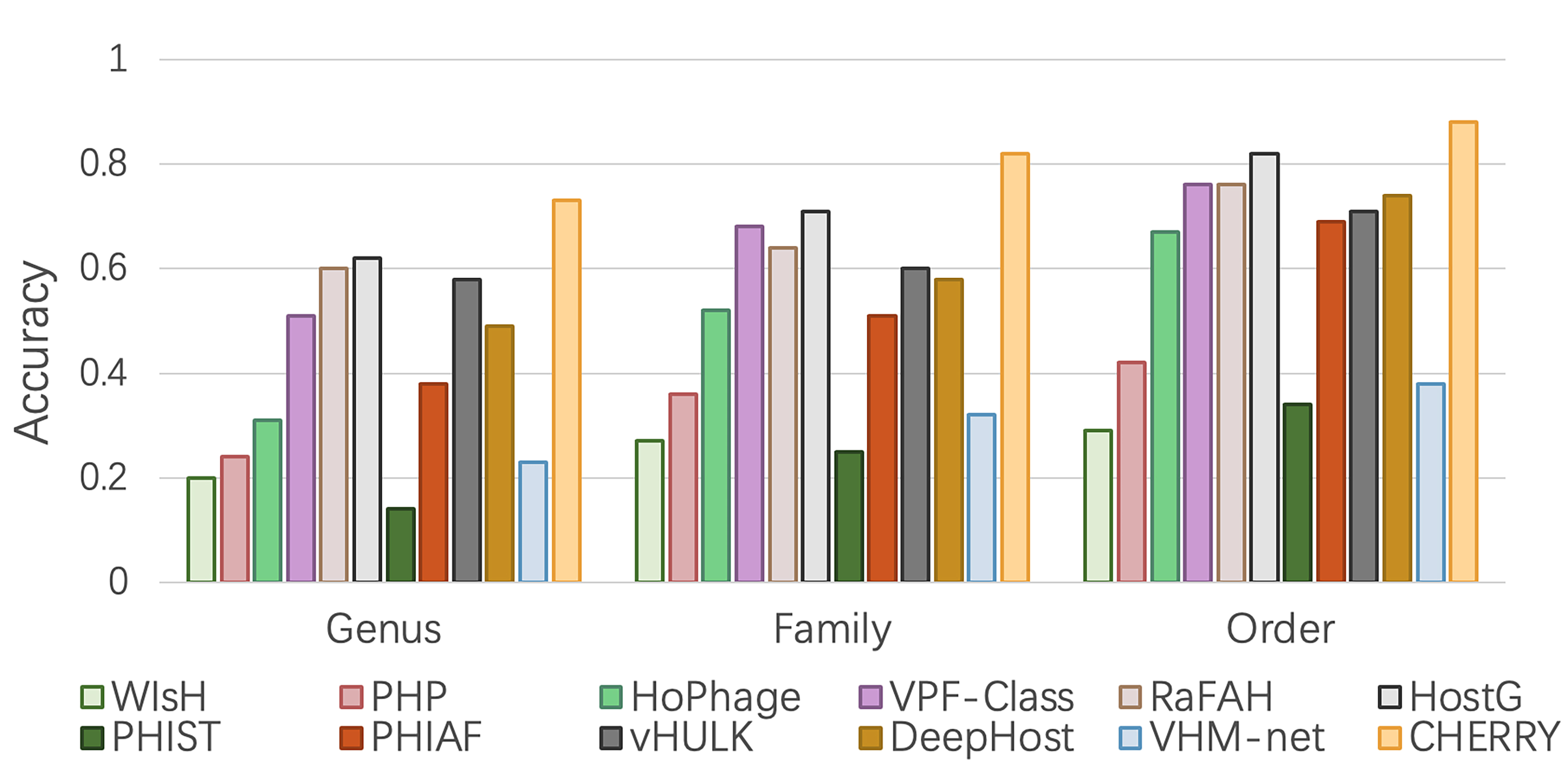}
    \caption{Host prediction accuracy for viruses that lack significant alignments against the prokaryotes. X-axis: taxonomic rankings. Y-axis: accuracy.}
    \label{fig:no_align}
\end{figure}

\subsubsection{Sequence similarity between viruses and prokaryotes}
Then, we further investigated the performance of these tools on predicting hosts for viruses that lack significant similarities with prokaryotic genomes. All the test viruses that do not have BLASTN alignments with the reference prokaryotes are used for evaluation. As shown in Fig. \ref{fig:no_align}, the accuracy of most methods decreases. Nevertheless, CHERRY still renders the best performance. vHULK, RaFAH, VPF-Class and HostG have better accuracy than VHM-net, PHIST, WIsH, and PHIAF probably because they mainly rely on the viral protein similarity for host prediction. The experimental results shown in Fig. \ref{fig:similarity} and Fig. \ref{fig:no_align} reveal that integrating the virus-virus and virus-prokaryote relationships in the multimodal graph enhances host prediction.

\begin{figure}[h!]
    \centering
    \includegraphics[width=0.55\linewidth]{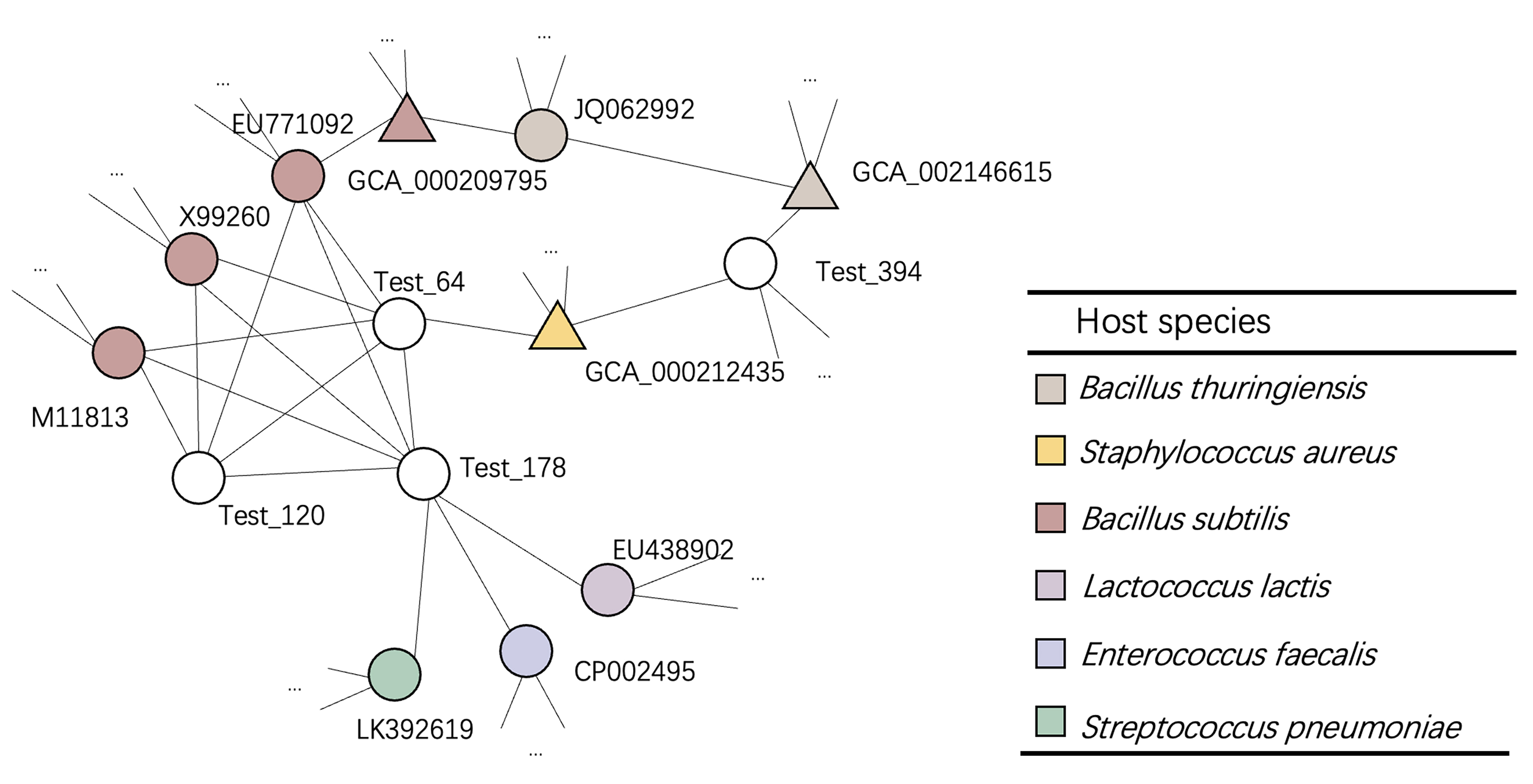}
    \caption{A case study for a sub-graph with heterogeneous labels. Triangles: prokaryotic nodes. Circles: virus nodes. White nodes: test viruses. Nodes with other colors: training samples. Different colors represent different species/labels. The open-end edges adjacent to the nodes indicate that these nodes have more connections.}
    \label{fig:case}
\end{figure}

\paragraph{A case study}
As described in Section \textit{Visualization of the knowledge graph}, we can have heterogeneous labels in sub-graphs.  An example is given in Fig. \ref{fig:case}. Test viruses (white nodes) connect with multiple nodes in different colors. Applying majority vote based on neighbors' labels will assign \textit{Bacillus subtilis} as the host for virsues `Test\_120', `Test\_64', and `Test\_178'. However, according to the given labels in the database, only `Test\_178' is a \textit{Bacillus subtilis} virus. `Test\_120' and `Test\_64' are \textit{Bacillus thuringiensis} viruses. 

The GCN-based encoder in CHERRY was not limited by the heterogeneous connections and predicted the hosts correctly. We recorded the prediction score of CHERRY in Table \ref{tab:aligncase}. We also reported the results of vHULK, VHM-net, PHIST, PHIAF, and DeepHost because they can predict hosts at the species level.

\begin{table}[!h]
\centering
\small
\begin{tabular}{p{2cm}p{2cm}p{3.4cm}}
\hline
\textbf{Method}                    & \textbf{Virus}     & \textbf{Prediction}             \\ \hline
\multirow{2}{*}{CHERRY}   & Test\_64  & \textbf{\textit{Bacillus thuringiensis}} \\
                          & Test\_120 & \textbf{\textit{Bacillus thuringiensis}} \\ \hline
\multirow{2}{*}{vHULK}    & Test\_64  & \textit{Bacillus subtilis}      \\
                          & Test\_120 & \textbf{\textit{Bacillus thuringiensis}} \\ \hline
\multirow{2}{*}{VHM-net}  & Test\_64  & \textit{Bacillus subtilis}      \\
                          & Test\_120 & \textit{Staphylococcus aureus}  \\ \hline
\multirow{2}{*}{PHIST}    & Test\_64  & \textit{Bacillus anthracis}     \\
                          & Test\_120 & -                      \\ \hline
\multirow{2}{*}{PHIAF}    & Test\_64  & \textit{Bacillus anthracis}     \\
                          & Test\_120 & \textit{Staphylococcus aureus}  \\ \hline
\multirow{2}{*}{DeepHost} & Test\_64  & \textit{Lactococcus lactis}     \\
                          & Test\_120 & \textit{Lactococcus lactis}     \\ \hline
\end{tabular}
\caption{Host predictions for two hard cases (Test\_120 and Test\_64 in Fig. \ref{fig:case}) using six methods that can predict the host at the species level. `-': no prediction. The correct prediction is in bold font.}
\label{tab:aligncase}
\end{table}

The results show that CHERRY can predict both viruses correctly and vHULK can assign host to one of the viruses correctly. Other methods failed to provide correct predictions. A plausible explanation is CHERRY not only considers alignment similarity/connections in the multimodal graph but also considers the $k$-mer frequency of the genomes when training the link prediction decoder. In summary, 73.7\% of viruses have multiple alignments/connections with different labelled nodes (viruses or prokaryotes) in our knowledge graph. While using only the graph topology might return ambiguous/wrong predictions, CHERRY can provide more accurate host identification for these viruses.

\subsection{Performance on short contigs}

Because the assembly programs may not generate complete virus genomes from viral metagenomic data, virus host prediction programs should be able to predict hosts on assembled contigs. To evaluate the robustness of CHERRY on short contigs, we generated DNA segments with different length ranges. 

\begin{figure}[h!]
    \centering
    \includegraphics[width=0.55\linewidth]{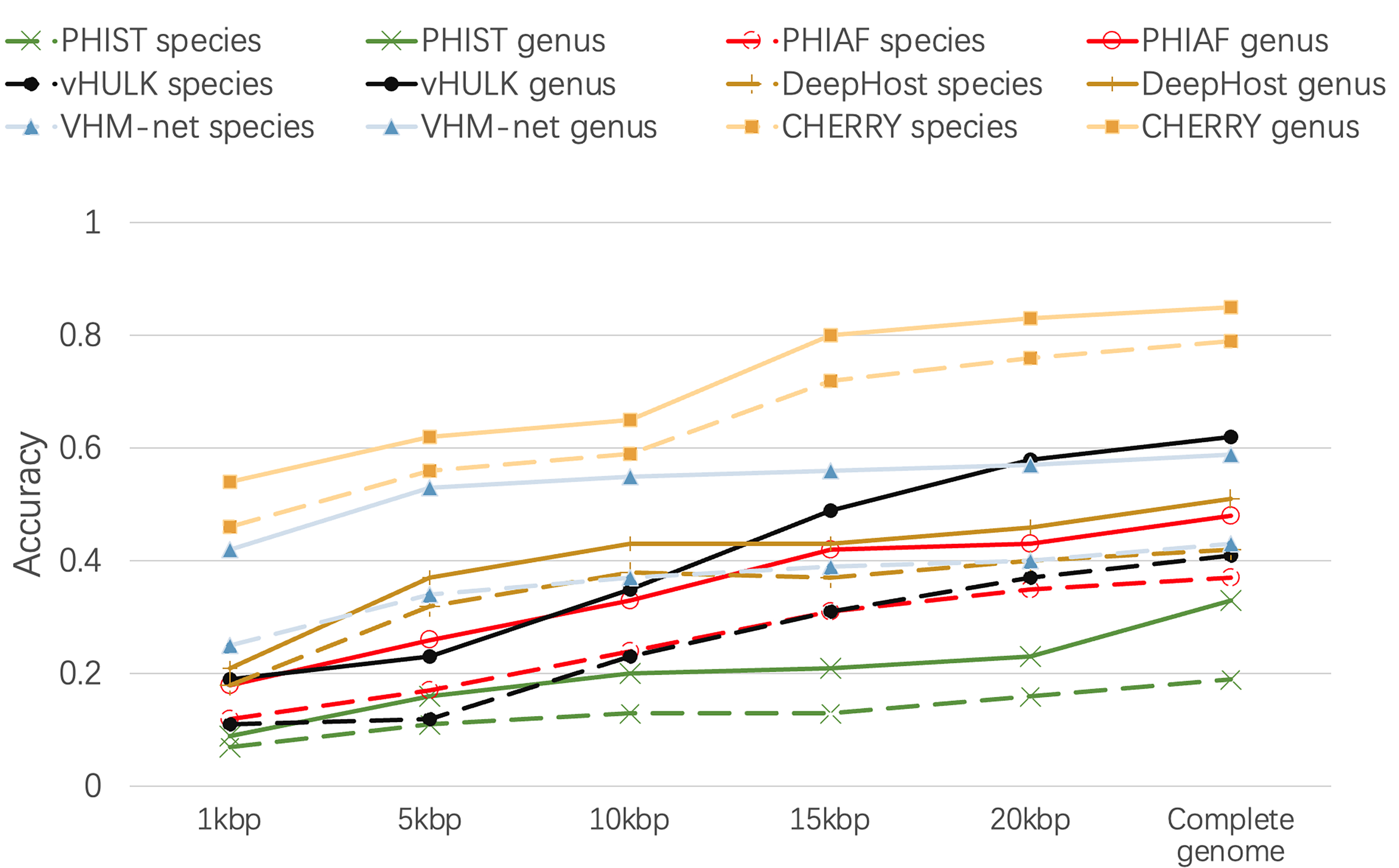}
    \caption{Host prediction performance on contigs. X-axis: length of the input contigs. Y-axis: accuracy.}
    \label{fig:short_contig}
\end{figure}

First, we generated contigs by cutting the test viruses' genomes with five different lengths: 1kbp, 5kbp, 10kbp, 15kbp, and 20kbp. We pick a random starting position in the genome and cut a substring of the given length. We repeated this process multiple times until we have sufficient contigs. Finally, we generated 6,340 contigs for each length. We used these contigs to evaluate PHIST, PHIAF, vHULK, DeepHost, VHM-net, and CHERRY, which can predict hosts at the species level, and reported the average accuracy in Fig. \ref{fig:short_contig}. As the figure shows, although the performance of all methods decreases with the decrease of the contigs’ length, CHERRY still achieves the best performance under all different lengths.

\subsection{Performance on different viral families}

Although CHERRY has the best performance in host prediction, its accuracy at the species level is slightly lower than 0.8. In order to identify the reasons, we conducted a closer examination of CHERRY's performance for different viral families. \textit{Caudovirales}, which contains phages with tails, is the order with the largest number of  sequenced prokaryotic viruses in the RefSeq database. The test set is dominated by three families under \textit{Caudovirales}: \textit{Siphoviridae}, \textit{Myoviridae}, and \textit{Podoviridae}. Thus, we group the phages by their family taxonomy and record the host prediction results accordingly.

\begin{figure}[h!]
    \centering
    \includegraphics[width=0.55\linewidth]{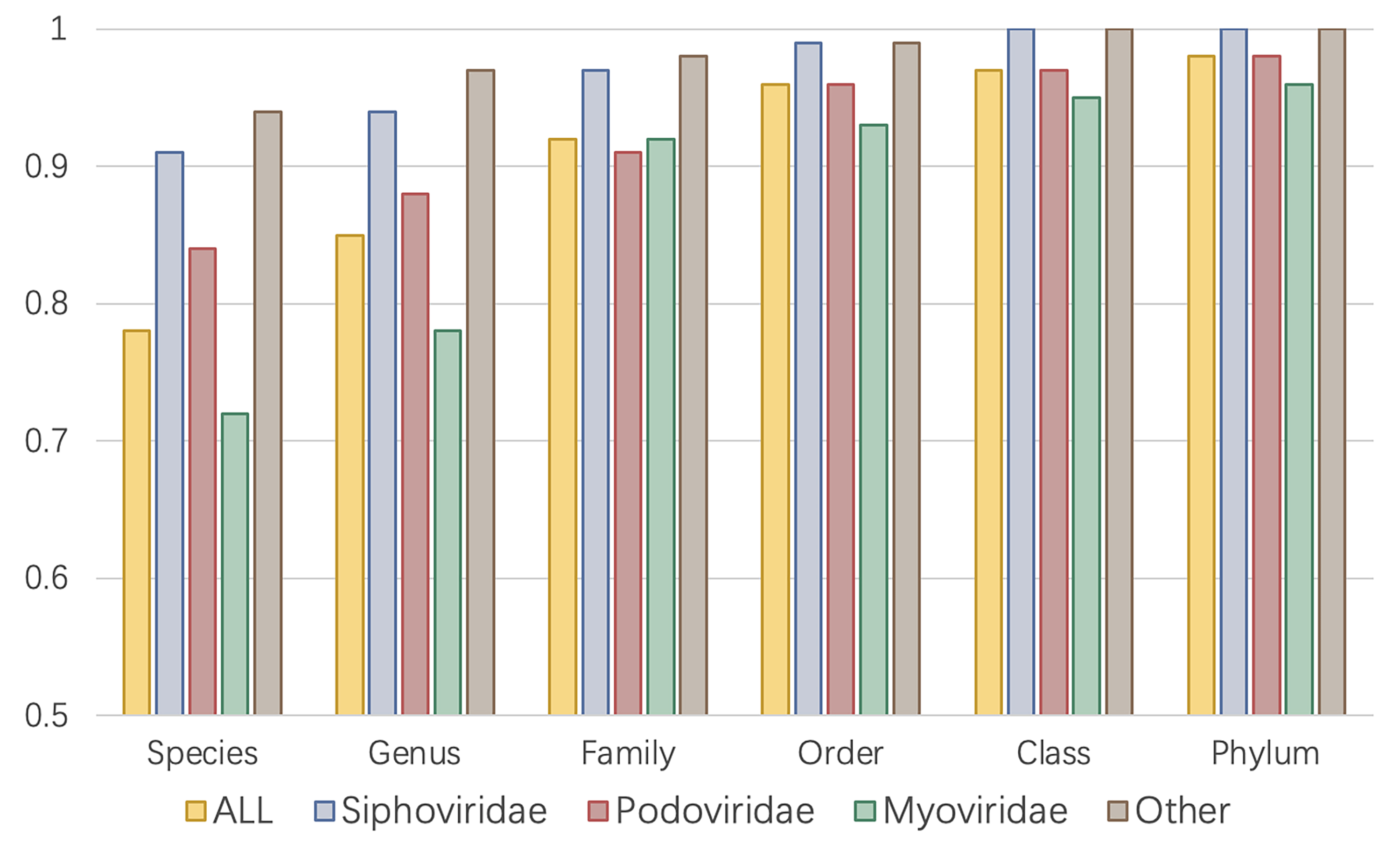}
    \caption{Host prediction results for different groups of viruses. X-axis: taxonomic rank. Y-axis: accuracy. ALL: the accuracy on the whole dataset, which is the same as Fig. \ref{fig:train_valid} (B). Other: accuracy for viruses that do not belong to \textit{Caudovirales}. }
    \label{fig:taxa}
\end{figure}

As shown in Fig. \ref{fig:taxa}, The accuracy of viruses that do not belong to \textit{Caudovirales} is always the best. The performance of phages in \textit{Mydoviridae} is worse than other groups at species and genus level but increases largely at the family level. One possible reason is that, as discussed in \cite{chibani2004phage, hamdi2017characterization}, some phages in \textit{Myoviridae} have the potential to infect multiple hosts from different species and even genera. But our data only contains one host species for each virus. Thus, for the viruses in \textit{Myoviridae}, some of the false predictions might indicate that the viruses of interest infect multiple hosts.

\begin{figure*}[h!]
    \centering
    \includegraphics[width=0.85\linewidth]{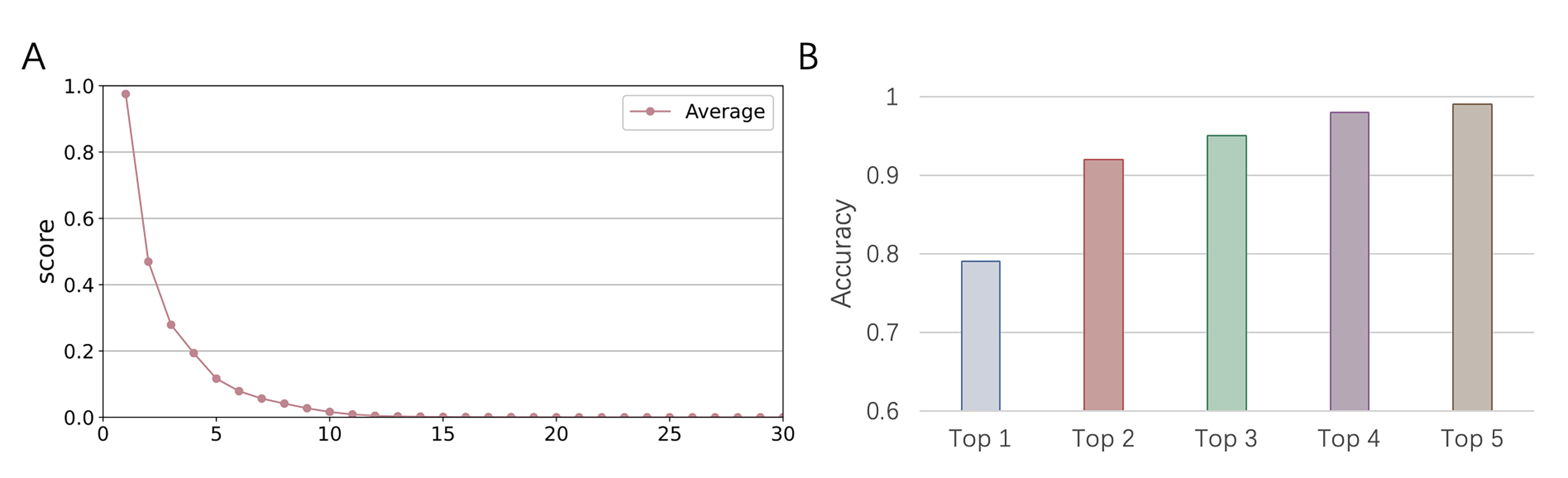}
    \caption{The experimental results of top-$k$ prediction. A: Tendency of the prediction score. X-axis: the sorted index by $k$. Y-axis: average score of the top-$k$ prediction. B: The accuracy using top-$k$ prediction. }
    \label{fig:topk}
\end{figure*}

\subsection{Top $k$ prediction scores}
Given the possibility of multiple hosts for some viruses, we also provide an alternative evaluation metric based on top $k$ predictions. Instead of only reporting the prokaryote with the highest prediction score, CHERRY also allows users to output multiple hosts based on ranked scores. 
In this section, we will first show the tendency of the prediction score. Because there are 60,105 candidate prokaryotic genomes for each testing virus, the decoder will score each virus-prokaryote pair, leading to 60,105 sorted scores for each test virus. We will calculate the average of the highest score for all test viruses, the second highest score, etc. Because there are 60,105 values, we only show the highest 30 scores in Fig. \ref{fig:topk}(A). The results show that the average score drops precipitously after the first ten values, suggesting that CHERRY has a strong selection preference for a few possible hosts. 

In Fig. \ref{fig:topk}(B), we present the top-$k$ host prediction accuracy using the 60,105-dimensional score vector. Specifically, if the real host label of a test virus exists in the highest $k$ predictions (scores), we will consider this as a correct prediction for the top-$k$ accuracy. As shown in Fig. \ref{fig:taxa}(B), the top-2 accuracy increases largely from top-1, and the growth trend becomes slower after that. We also found that even if the scores of some real virus-host interactions are not the top 1 score, their scores are usually larger than 0.9, indicating that they are predicted with high confidence. Thus, CHERRY can support a method for further improving the host prediction results: using score threshold (0.9) and outputting top k predictions above the threshold. Although this method might predict more than one virus-prokaryote interaction for a given virus, it can largely narrow the search scope of the potential host. 

\subsection{Host prediction on metagenomic data}
In this section, we will validate CHERRY on host prediction for possibly novel viruses from metagenomic data. Because metagenomic data usually contains many different species or components, such as eukaryotic viruses or plasmids, users should run prokaryotic virus identification tools to screen viral contigs from metagenomic data before using CHERRY for host prediction. For example, Metaviral spades \cite{antipov2020metaviral}, Seeker \cite{auslander2020seeker}, and VirSorter \cite{roux2015virsorter} are widely used for virus identification. We first evaluate CHERRY on a Hi-C sequencing dataset that provides annotated phage-host interactions \cite{marbouty2021metahic}.  Then, we choose two newly published metagenomic datasets containing viruses in two habitats: glacier \cite{zhong2021glacier} and human gut \cite{benler2021thousands, marbouty2021metahic}. The authors used assembly tools and virus identification tools, such as VirSorter \cite{roux2015virsorter}, to obtain virus-originating contigs from the samples. Then, we applied CHERRY to predict hosts for these virus contigs.

\subsubsection{Case study one: phage-host interactions derived using MetaHiC}

Recently, a metagenomic Hi-C approach named MetaHiC was applied to human gut samples to identify interactions between phages and assembled bacterial genomes \cite{marbouty2021metahic}. Based on the design of MetaHiC, this approach can capture DNA-DNA collisions during phages' replication inside bacterial cells, thus providing a quality test set for computational host prediction. In this experiment, we use the phage-bacteria interactions provided by the this study \cite{marbouty2021metahic} as the ground truth to evaluate the performance of the state-of-the-art methods.

\begin{figure*}[h!]
    \centering
    \includegraphics[width=0.65\linewidth]{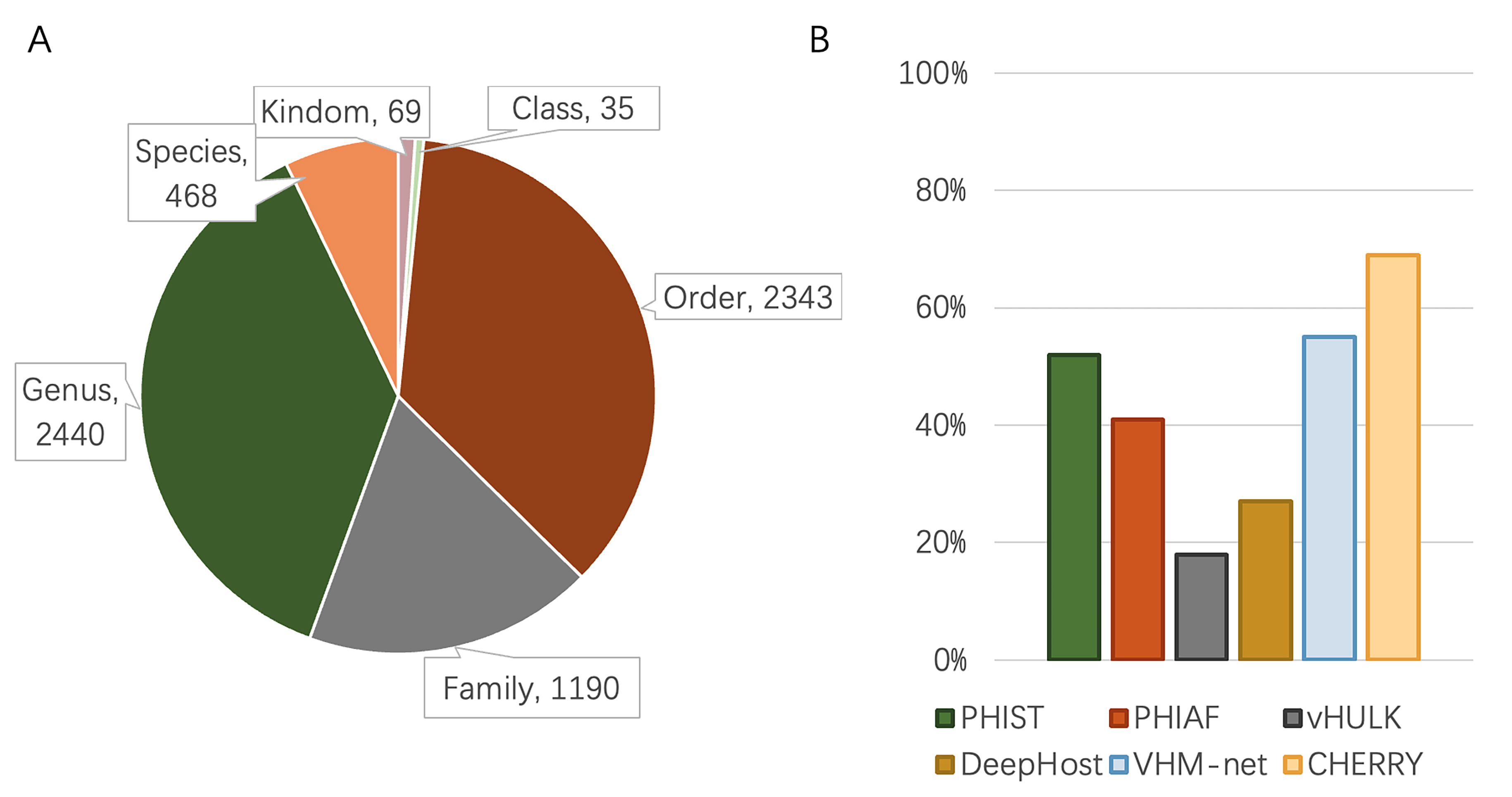}
    \caption{The experimental results on the MetaHiC dataset. For each bin, we use the lowest rank of the assigned taxon as the host label for the phage contigs in the bin. Phage contigs from the same bins have the same label. A: The number of phage contigs with host labels at different taxonomic ranks. B: The host prediction accuracy (Y-axis) on the 6,545 phage contigs. The comparison includes six tools that can predict hosts at species level.}
    \label{fig:hic}
\end{figure*}

We downloaded the supplementary data from the paper’s GitHub repository. The authors provided the taxonomic annotations for contigs and bins in these data files, enabling us to extract the phage contigs from each bin and label the phage contigs’ hosts using the bin’s taxon. Finally, we have 995 bins and 6,545 phage contigs. We then removed the phage contigs from the bins, and thus, all these phage contigs can be used as the test set. The length of these phage contigs ranges from 1kbp to 10kbp. Because each bacterial bin is assigned to a taxon that ranges from kingdom to species, we used the lowest rank as the host label for each phage contig. For example, if the lowest rank of a bin's annotation is a family, the host label of the phage contig is the name of this family. The distribution of phage contigs with different ranks of host labels is summarized in Fig. \ref{fig:hic} (A). If a tool can predict the same label as the given one (the lowest rank) by MetaHiC, the prediction will be counted as correct. Out of the 12 tools we tested in this work, only PHIST, PHIAF, vHULK, DeepHost, VHM-net, and CHERRY can predict hosts at the species level; thus we compared their accuracy in Fig. \ref{fig:hic} (B). CHERRY still has the highest accuracy. PHIST also shows good performance for predicting new interactions. vHULK and DeepHost have lower accuracy, showing that they are not optimized for predicting new interactions.  


\subsubsection{Case study two: newly identified viruses in glacier metagenomic data}

This data set was sequenced from the core of the glacier \cite{zhong2021glacier}. Due to global warming, the melting glacier might release those ancient viruses to the environment in the future. Metagenomic sequencing provides a powerful means to study the virus composition.
The authors reported 33 virus contigs identified by VirSorter \cite{roux2015virsorter} with high confidence. The length of the contigs range from 12,041 to 93,811. According to the authors' analysis, these metagenomic data contains four dominant bacterial genera including \textit{Methylobacterium}, \textit{Sphingomonas}, \textit{Janthinobacterium}, and \textit{Herminiimonas} and 3 putative laboratory contaminants including \textit{Synechococcus phages}, \textit{Cellulophaga phages}, and \textit{Pseudoalteromonas phages}. Thus, we use the bacteria under these genera with all prokaryotes in our database as candidate hosts and run CHERRY. 

\begin{figure}[h!]
    \centering
    \includegraphics[width=0.3\linewidth]{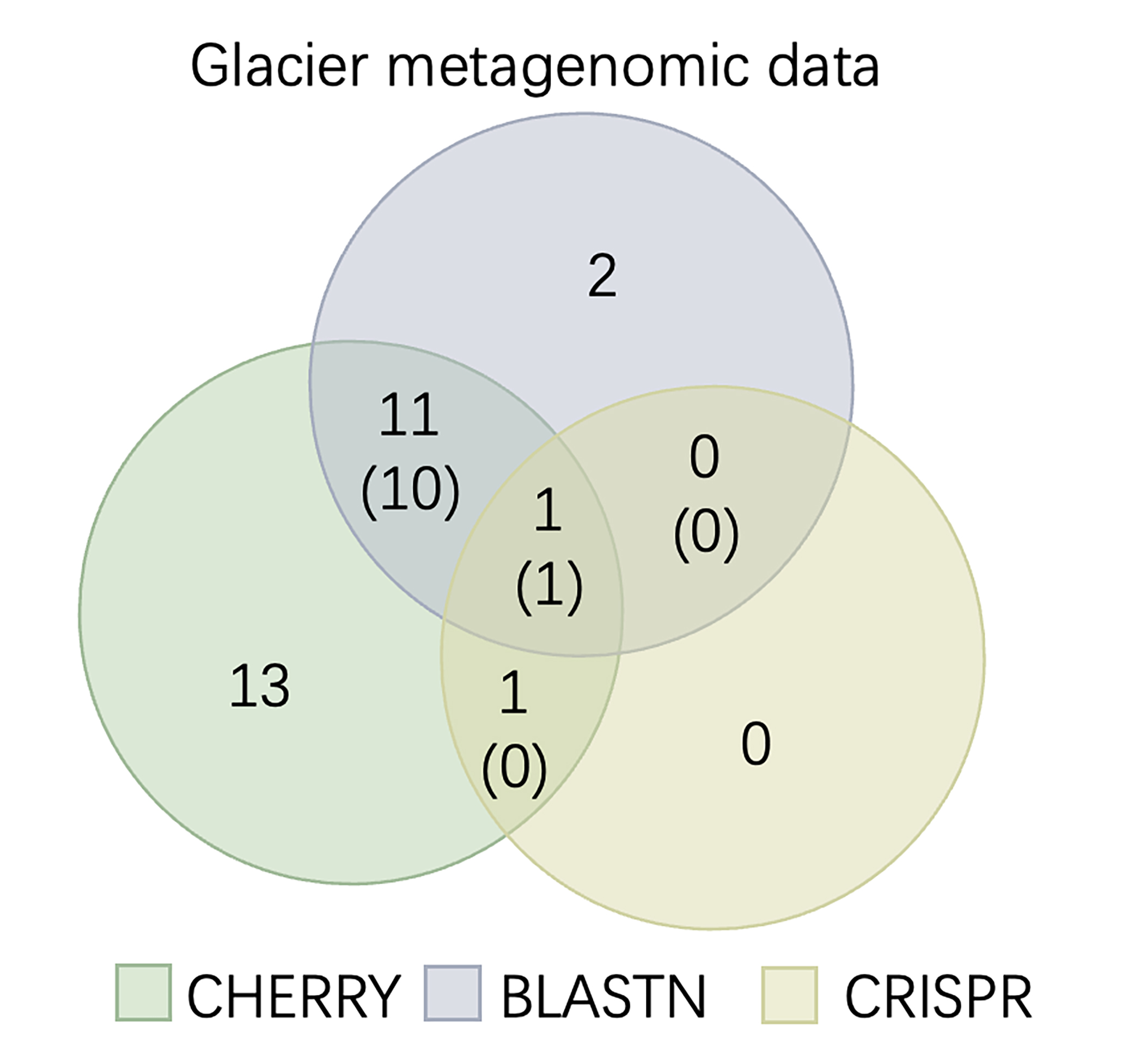}
    \caption{Host prediction on the glacier metagenomic data. The numbers without parentheses represent the number of viruses. The numbers with parentheses represent the number of viruses with the same predicted hosts. For example, 12 viruses have predictions by both CHERRY and BLASTN and  11 of them have the same predicted hosts.}
    \label{fig:glacier}
\end{figure}

Because the authors already reported the host prediction results using BLASTN and CRISPR, we re-used their predictions and compared them with our method. To report a more precise prediction, we only keep the predictions of prokaryotes with a score larger than 0.9. Thus, some viruses do not have predicted hosts. The Venn diagram of the three methods is shown in Fig. \ref{fig:glacier}. 
CHERRY can predict hosts for more viruses (about 80\% of 33 contigs) on this dataset. This is expected because our previous analysis and experiments have shown that BLASTN and CRISPR can only predict hosts for a very limited number of viruses. For the viruses with predictions from either BLASTN or CRISPR, the prediction of CHERRY is largely consistent with them. Specifically, among the 14 BLASTN-predicted viruses and 2 CRISPR-predicted viruses, CHERRY has 11 identical predictions as BLASTN and one identical prediction as CRISPR.

\subsubsection{Case study three: newly identified viruses in gut metagenomic data}

In a newly published human gut metagenomic study \cite{benler2021thousands}, the authors identified 3,738 complete viruses genomes that represent 451 putative genera using both ViralVerify \cite{antipov2020metaviral} and Seeker \cite{auslander2020seeker}. Investigating the hosts of these viruses will shed lights on both the composition and functional analysis of the underlying metagenomic data.

\begin{figure}[h!]
    \centering
    \includegraphics[width=0.3\linewidth]{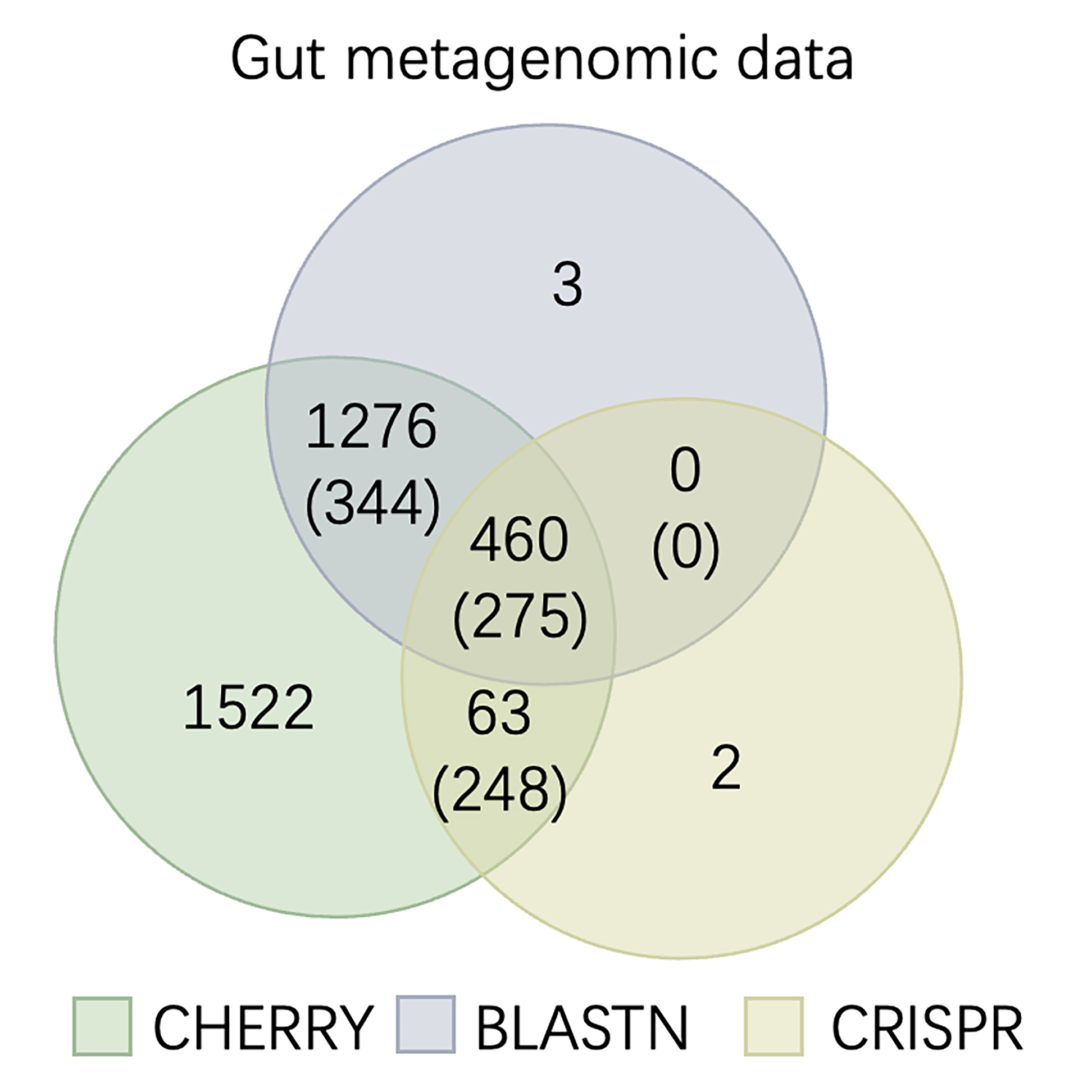}
    \caption{Host prediction on the gut metagenomic data. The numbers without parentheses represent the number of viruses. The numbers with parentheses represent the number of viruses with the same predicted hosts.}
    \label{fig:gut}
\end{figure}

We reused the reported results of CRISPR from \cite{benler2021thousands} and ran BLASTN and CHERRY. Because there is no prior information of the prokaryotes in the this metagenomic data, we use all prokaryotes in our database to construct a graph. The result is shown in Fig. \ref{fig:gut}. CHERRY significantly improves the number of predicted viruses compared to BLASTN and CRISPR. About 89\% (3,321/3,738) of viruses were assigned hosts by CHERRY with a score threshold of 0.9. What's more, the predictions of CHERRY are highly consistent with CRISPR. Only two virus contigs predicted by CRISPR have no predictions by CHERRY because their scores are less than the threshold. All other CRISPR-predicted contigs (460+63) have the same labels (275+248) as CHERRY. We also found that although BLASTN output hosts for 2,131 (57\%) viruses, many have multiple alignments. Only 59\% (275/460) of the BLAST predictions are consistent with CRISPR. Considering that CRISPR is a reliable signal for host prediction, CHERRY's output is consistent with our previous experiments, demonstrating both high high accuracy. 

\subsection{Predicting viruses that infect prokaryotes}
CHERRY can estimate how likely there is a link (infection) between a virus $p_i$ and a prokaryote $h_i$. Thus, CHERRY can also be used to output viruses that infect a prokaryote of interest. We use Eq. \ref{m2} to predict the viruses that infect given prokaryotic genomes.  CHERRY will take prokaryotic genomes as input and output viruses with prediction scores above a given threshold. This function can help users identify candidate viruses that can infect targeted prokaryotes.
We use $recall$ and $precision$ introduced in the \textit{Experimental setup} Section to evaluate the performance. 

\begin{figure}[h!]
    \centering
    \includegraphics[width=0.45\linewidth]{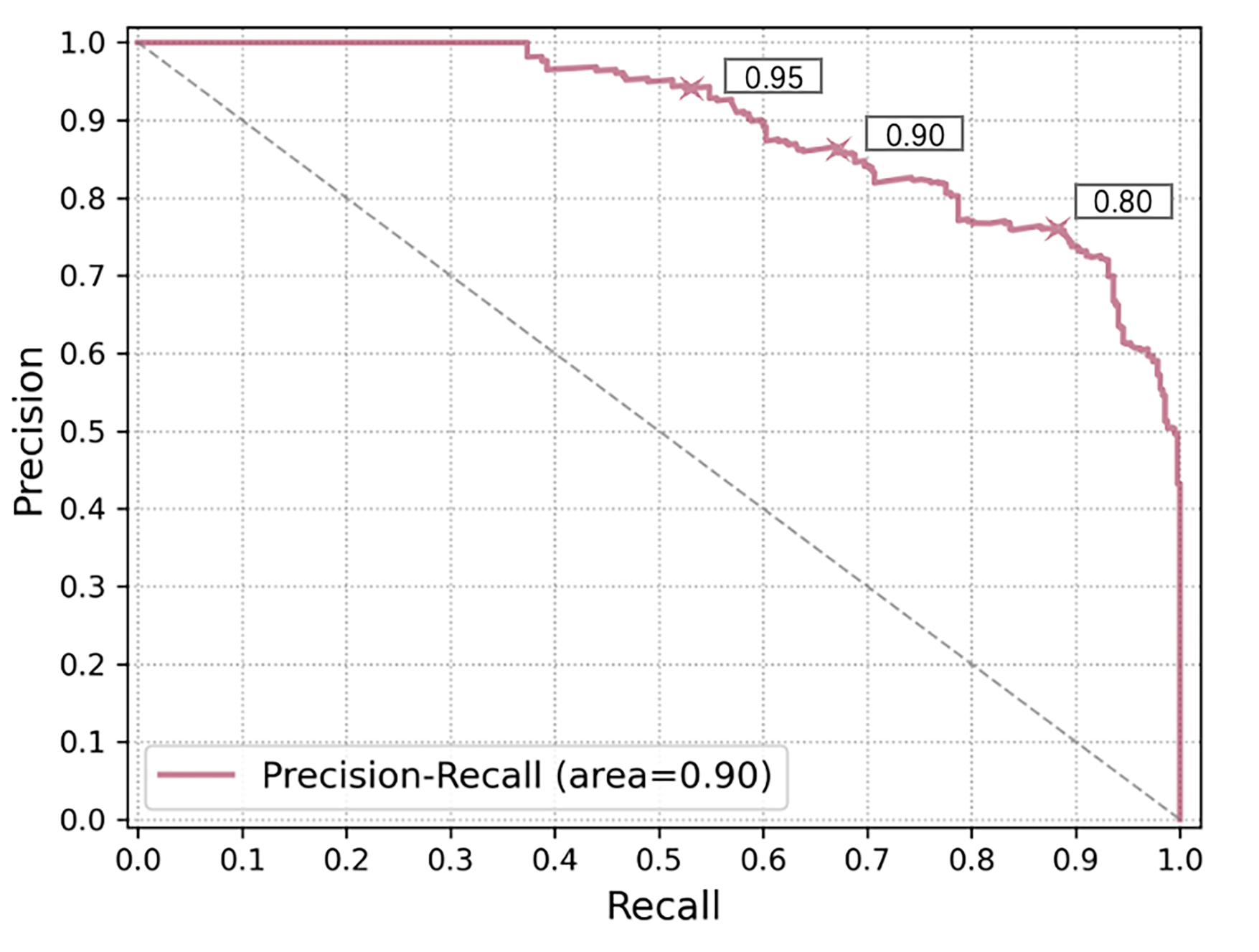}
    \caption{The precision-recall curve of predicting viruses infecting targeted prokaryotes. X-axis: recall, Y-axis: precision. The performance for three thresholds 0.95, 0.9, and 0.8 are marked with the the cross sign on the curve.}
    \label{fig:curve}
\end{figure}

As shown in Fig. \ref{fig:curve}, we draw the precision-recall curve by recording the precision and recall under different thresholds. When using a more lenient threshold, the recall increases with a sacrifice of the precision. Users can choose the thresholds according to their needs. In order to achieve high precision, we use 0.9 as the default threshold in our program.

\section{Discussion}
In this work, we describe a new virus host prediction tool, which formulates the host prediction problem as link prediction in a multimodal graph. 
This multimodal graph integrates different types of prior knowledge, including protein organization, CRISPR, sequence similarity, and $k$-mer frequency, from both labeled (training) and unlabeled (test) data. Then we apply a graph convolutional encoder to embed feature vectors on the graph and use a 2-layer neural network decoder to calculate the probability of a query (virus-prokaryote) pair forms an interaction. We apply the end-to-end training process and the negative sampling method to optimize the loss. This semi-supervised learning scheme helps the model learn features from both the training set and test set and thus leads to high prediction accuracy. The large-scale experiments on 1,940 viruses and 60,105 prokaryotic genomes show that we improve the host prediction accuracy from 43\% to \textasciitilde80\% at the species level. We also use two case studies to validate the reliability and practicality of our model in real-world applications. 

Although CHERRY has greatly improved the performance of host prediction, we will further optimize or extend CHERRY in our future work.First, CHERRY currently only uses sequence-based features such as sequence similarity and $k$-mer frequency. 
Because the binding between receptor-binding proteins (RBP) and the host cells' receptors is essential for the virus to gain entry into the host cells, one possible extension is to include protein-protein interactions (PPI) between RBPs and receptors in the edge construction. However, because only a few PPIs about prokaryotic viruses are reported, more experiments or computational structure and interaction predictions are needed to augment the graph. Second, other sequence-based features can be added to the node features if those features can help host prediction. We will support users to add their customized features to enhance the learning ability. Third, CHERRY is trained for species-level host prediction and is not optimized for strain-level host prediction. The high similarity between strains can lead to ambiguous predictions. In addition, another challenge is the fewer training samples at the strain level. We will explore whether CHERRY can be extended for strain-level host prediction in our future work. Finally, our method can benefit from using more characterized virus-host interactions. In order to investigate whether CHERRY can be generalized to new viruses, the current version of CHERRY was trained using viruses submitted before 2015. If we use viruses submitted before 2014 as the training set, the decreased number of labeled nodes (from 1306 to 968) reduces the species-level accuracy from 78\% to 73\% on the test set. Thus, with increased availability of characterized interactions, the accuracy of CHERRY will increase too.

\section{Data Availability}
All data and codes used for this study are available online or upon request to the authors. The source code of CHERRY is available via: \url{https://github.com/KennthShang/CHERRY}. The accessions of training set and test set are available via: \url{https://github.com/KennthShang/CHERRY/Interactiondata}. The training set is listed in VHM\_PAIR\_TAX.xls. The test set is listed in TEST\_PAIR\_TAX.xls.

\section{Conflict of Interest}
There is no competing interest.



\section{Funding}
City University of Hong Kong (Project 9678241),  HKIDS (9360163), and the Hong Kong Innovation and Technology Commission (InnoHK Project CIMDA).

\bibliographystyle{unsrt}  
\bibliography{references}  

\end{document}